\documentclass[journal,11pt,draftcls,onecolumn]{IEEEtran}%
\usepackage{amsfonts}
\usepackage{amsmath}
\usepackage{amssymb}
\usepackage{graphicx}%
\newtheorem{theorem}{Theorem}

\newtheorem{corollary}{Corollary}

\newtheorem{definition}{Definition}

\newtheorem{remark}{Remark}

\begin{document}

\title{Ergodic Fading Interference Channels:\ Sum-Capacity and Separability}
\author{Lalitha Sankar,~\IEEEmembership{Member,~IEEE,} Xiaohu
Shang,~\IEEEmembership{Member,~IEEE,} Elza
Erkip,~\IEEEmembership{Senior~Member,~IEEE,} and~H.\ Vincent
Poor,~\IEEEmembership{Fellow,~IEEE}\thanks{L. Sankar, X. Shang, and H. V. Poor
are with the Department of Electrical\ Engineering, Princeton University,
Princeton, NJ 08544, USA. email: \{lalitha,xshang,poor@princeton.edu\}. E.
Erkip is with the Department of Electrical and Computer Engineering,
Polytechnic Institute of New York University, Brooklyn, NY 11201, USA. email:
elza@poly.edu. This research was conducted in part when E.\ Erkip was visiting
Princeton University.}\thanks{This research was supported in part by the
National\ Science Foundation under Grant CNS-06-25637 and in part by a
fellowship from the Princeton University Council on\ Science and Technology.
The material in this paper was presented in part at the IEEE\ International
Symposium on Information Theory, Toronto, Canada, Jul. 2008 and at the
$46^{th}$ Annual Allerton Conference on Communications, Control, and
Computing, Monticello, IL, Sep. 2008.}}
\pubid{~~}
\maketitle

\begin{abstract}
The sum-capacity of ergodic fading Gaussian two-user interference channels
(IFCs) is developed under the assumption of perfect channel state information
at all transmitters and receivers. For the sub-classes of \textit{uniformly
strong }(every fading state is strong) and \textit{ergodic very strong}
two-sided IFCs (a mix of strong and weak fading states satisfying specific
fading averaged conditions) the optimality of completely decoding the
interference, i.e., converting the IFC to a compound multiple access
channel\ (C-MAC), is proved. It is also shown that this capacity-achieving
scheme requires encoding and decoding jointly across all fading states. As an
achievable scheme and also as a topic of independent interest, the capacity
region and the corresponding optimal power policies for an ergodic fading
C-MAC are developed. For the sub-class of \textit{uniformly weak} IFCs (every
fading state is weak), genie-aided outer bounds are developed. The bounds are
shown to be achieved by ignoring interference and separable coding for
one-sided fading IFCs. Finally, for the sub-class of one-sided \textit{hybrid}
IFCs (a mix of weak and strong states that do not satisfy ergodic very strong
conditions), an achievable scheme involving rate splitting and joint coding
across all fading states is developed and is shown to perform at least as well
as a separable coding scheme.

\end{abstract}


\IEEEpeerreviewmaketitle

\section{Introduction}

The interference channel (IFC) models a wireless network where every
transmitter (user) communicates with its unique intended receiver while
causing interference to the remaining receivers. For the two-user IFC, the
topic of study in this paper and henceforth simply referred to as an IFC, the
capacity region is not known in general even when the channel is
time-invariant, i.e., non-fading. Capacity results are known only for specific
classes of non-fading two-user IFCs where the classes are identified by the
relative strength of the channel gains of the interfering cross-links and the
intended direct links. Thus, strong and weak IFCs refer to the cases where the
channel gains of the cross-links are at least as large as those of the direct
links and vice-versa.

The capacity region for the class of strong Gaussian IFCs is developed
independently in
\cite{cap_theorems:Sato_IC,cap_theorems:Carleial_VSIFC,cap_theorems:KobayashiHan_IC}
and can be achieved when both receivers decode both the intended and
interfering messages. In contrast, for the weak channels, the sum-capacity can
be achieved by ignoring interference when the channel gains of one of the
cross-links is zero, i.e., for a one-sided IFC \cite{cap_theorems:Costa_IC}.
More recently, the sum-capacity of a class of noisy or very weak Gaussian IFCs
has been determined independently in \cite{cap_theorems:ShangKramerChen},
\cite{cap_theorems:MotaKhan}, and \cite{cap_theorems:AR_VVV}. Outer bounds for
the IFC are developed in \cite{cap_theorems:Kramer_IFCOB} and
\cite{cap_theorems:ETW} while several achievable rate regions for the Gaussian
IFC are studied in \cite{cap_theorems:I_Sason_IFC}.

The best known inner bound is due to Han and Kobayashi (HK)
\cite{cap_theorems:KobayashiHan_IC}. Recently, in \cite{cap_theorems:ETW} a
simple HK type scheme is shown to achieve every rate pair within 1 bit/s/Hz of
the capacity region. In \cite{cap_theorems:Weng_Tuninetti}, the authors
reformulate the HK region as a sum of two sets to characterize the maximum
sum-rate achieved by Gaussian inputs and without time-sharing. More recently,
the approximate capacity of two-user Gaussian IFCs is characterized using a
deterministic channel model in \cite{cap_theorems:Bresler_Tse}. The
sum-capacity of the class of non-fading MIMO IFCs is studied in
\cite{cap_theorems:Shang_MIMOIFC}.

Relatively fewer results are known for parallel or fading IFCs. In
\cite{cap_theorems:ChuCioffi_IC}, the authors develop an achievable scheme of
a class of two-user parallel Gaussian IFCs where each parallel channel is
strong using independent encoding and decoding in each parallel channel. In
\cite{cap_theorems:SumCap_ParZIFC}, Sung \textit{et al.} present an achievable
scheme for a class of one-sided two-user parallel Gaussian IFCs. The
achievable scheme involves encoding and decoding signals over each parallel
channel independently such that, depending on whether a parallel channel is
weak or strong (including very strong) one-sided IFC, the interference in that
channel is either viewed as noise or completely decoded, respectively. In this
paper, we show that independent coding across sub-channels is in general not
sum-capacity optimal.

Recently, for parallel Gaussian IFCs, \cite{cap_theorems:Shang_03} determines
the conditions on the channel coefficients and power constraints for which
independent transmission across sub-channels and treating interference as
noise is optimal. Techniques for MIMO IFCs \cite{cap_theorems:Shang_MIMOIFC}
are applied to study separability in parallel Gaussian\ IFCs (PGICs) in
\cite{cap_theorems:KaistParIFC}. It is worth noting that PGICs are a special
case of ergodic fading IFCs in which each sub-channel is assigned the same
weight, i.e., occurs with the same probability; furthermore, they can also be
viewed as a special case of MIMO IFCs and thus results from MIMO IFCs can be
directly applied.

For fading interference networks with three or more users, in
\cite{cap_theorems:CadamJafar_IFCAlign}, the authors develop an
\textit{interference alignment} coding scheme to show that the sum-capacity of
a $K$-user IFC scales linearly with $K$ in the high signal-to-noise ratio
(SNR) regime when all links in the network have similar channel statistics.

\begin{figure}[tbp] \centering
{\includegraphics[
height=2.6705in,
width=5.5867in
]%
{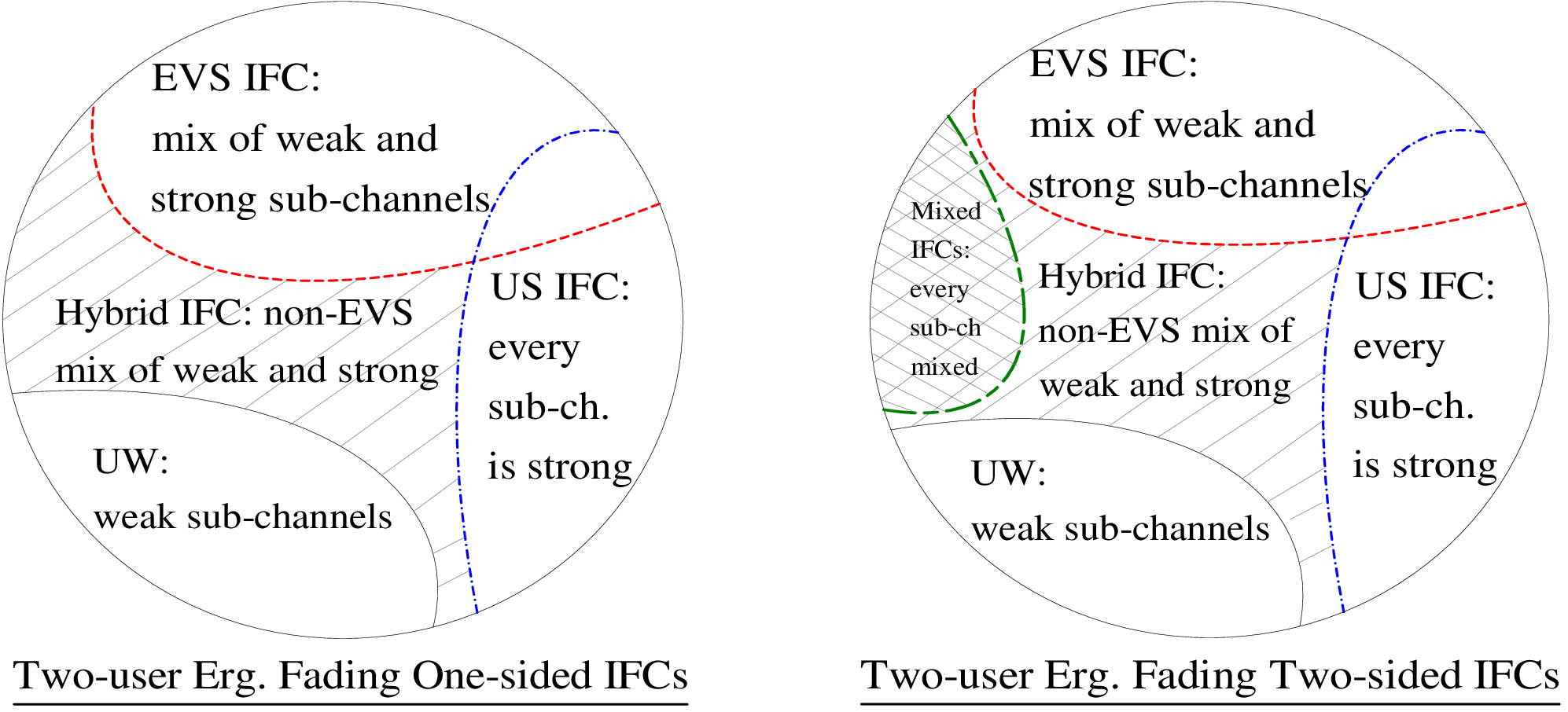}%
}%
\caption{A Venn diagram representation of the four sub-classes of ergodic fading one- and two-sided IFCs.}\label{FigIFCVenn}%
\end{figure}

In this paper, we study ergodic fading two-user Gaussian IFCs and determine
the sum-capacity and the corresponding optimal power policies for specific
sub-classes, where we define each sub-class by the fading statistics. Noting
that ergodic fading IFCs are a weighted collection of parallel IFCs
(sub-channels), we identify four sub-classes that jointly contain the set of
all ergodic fading IFCs. We develop the sum-capacity for two of them. For the
third sub-class, we develop the sum-capacity when only one of the two
receivers is affected by interference, i.e., for a one-sided ergodic fading
IFC. While the four sub-classes are formally defined in the sequel, we refer
the reader to Fig. \ref{FigIFCVenn} for a pictorial representation. An
overview of the capacity results is illustrated in the sequel in Fig.
\ref{Fig_IFCVenn}.

A natural question that arises in studying ergodic fading and parallel
channels is the optimality of \textit{separable coding}, i.e., whether
encoding and decoding independently on each sub-channel is optimal in
achieving one or more points on the boundary of the capacity region. For each
sub-class of IFCs we consider, we address the optimality of separable coding,
often referred to as \textit{separability}, and demonstrate that in contrast
to point-to-point, multiple-access, and broadcast channels without common
messages
\cite{cap_theorems:GoldsmithVaraiya,cap_theorems:TH01,cap_theorems:Tse_BC},
separable coding is not necessarily sum-capacity optimal for ergodic fading IFCs.

The first of the four sub-classes is the set of \textit{ergodic very strong}
(EVS) IFCs in which each sub-channel can be either weak or strong but averaged
over all fading states (sub-channels) the interference at each receiver is
sufficiently strong that the two direct links from each transmitter to its
intended receiver are the bottle-necks limiting the sum-rate. For this
sub-class, we show that requiring both receivers to decode the signals from
both transmitters is optimal, i.e., the ergodic very strong IFC modifies to a
two-user ergodic fading compound multiple-access channel (C-MAC) in which the
transmitted signal from each user is intended for both receivers
\cite{cap_theorems:SEP}. To this end, as an achievable rate region for IFCs
and as a problem of independent interest, we develop the capacity region and
the optimal power policies that achieve them for ergodic fading C-MACs (see
also \cite{cap_theorems:SEP}).

For EVS IFCs we also show that achieving the sum-capacity (and the capacity
region) requires transmitting information (encoding and decoding) jointly
across all sub-channels, i.e., separable coding in each sub-channel is
strictly sub-optimal. Intuitively, the reason for joint coding across channels
lies in the fact that, analogous to parallel broadcast channels with common
messages \cite{cap_theorems:JindalGold}, both transmitters in the EVS IFCs
transmit only common messages intended for both receivers for which
independent coding across sub-channels becomes strictly sub-optimal. To the
best of our knowledge this is the first capacity result for fading two-user
IFCs with a mix of weak and strong sub-channels. For such mixed ergodic IFCs,
recently, a strategy of \textit{ergodic interference alignment} is proposed in
\cite{cap_theorems:Nazer01}, and is shown to achieve the sum-capacity in
\cite{cap_theorems:Jafar_ErgIFC} for a class of $K$-user fading IFCs with
uniformly distributed phase and at least $K/2$ disjoint equal strength
interference links.

The second sub-class is the set of \textit{uniformly strong }(\textit{US}%
)\textit{ }IFCs in which every sub-channel is strong, i.e., the cross-links
have larger fading gains than the direct links for each fading realization.
For this sub-class, we show that the capacity region is the same as that of an
ergodic fading C-MAC with the same fading statistics and that achieving this
region requires joint coding across all sub-channels.

The third sub-class is the set of \textit{uniformly weak }(\textit{UW}%
)\textit{ }IFCs for which every sub-channel is weak. As a first step, we study
the one-sided uniformly weak IFC and develop genie-aided outer bounds. We show
that the bounds are tight when the interfering receiver ignores the weak
interference in every sub-channel. Furthermore, we show that separable coding
is optimal for this sub-class. The sum-capacity results for the one-sided
channel are used to develop outer bounds for the two-sided case; however,
sum-capacity results for the two-sided case will require techniques such as
those developed in \cite{cap_theorems:Shang_03} that also determine the
channel statistics and power policies for which ignoring interference and
separable coding is optimal.

The final sub-class is the set of \textit{hybrid }IFCs for which the
sub-channels are a mix of strong and weak such that there is at least one weak
and one strong sub-channel but are not EVS IFCs (and by definition also not US
and UW\ IFCs). The capacity-achieving strategy for EVS and US IFCs suggest
that a joint coding strategy across the sub-channels can potentially take
advantage of the strong states to partially eliminate interference. To this
end, for ergodic fading \textit{one-sided IFCs}, we propose a general joint
coding strategy that uses rate-splitting and Gaussian codebooks without
time-sharing for all sub-class of IFCs. For two-sided IFCs, the coding
strategy we present generalizes to a two-sided HK-based scheme with Gaussian
codebooks and no time-sharing that is presented and studied in
\cite{cap_theorems:Tuninetti}.

In the non-fading case, a one-sided non-fading IFC is either weak or strong
and the sum-capacity is known in both cases. In fact, for the weak case the
sum-capacity is achieved by ignoring the interference and for the strong case
it is achieved by decoding the interference at the receiver subject to the
interference. However, for ergodic fading one-sided IFCs, in addition to the
UW\ and US sub-classes, we also have to contend with the hybrid and EVS
sub-classes each of which has a unique mix of weak and strong sub-channels.
The HK-based achievable strategy we propose applies to all sub-classes of
one-sided IFCs and includes the capacity-achieving strategies for the EVS, US,
and UW as special cases.

The sub-class of \textit{uniformly mixed }(\textit{UM})\textit{ }IFCs obtained
by overlapping two complementary one-sided IFCs, one of which is uniformly
strong and the other uniformly weak, belongs to the sub-class of hybrid
(two-sided) IFCs. For UM\ IFCs, we show that to achieve sum-capacity the
transmitter that interferes strongly transmits a common message across all
sub-channels while the weakly interfering transmitter transmits a private
message across all sub-channels. The two different interfering links however
require joint encoding and decoding across all sub-channels to ensure optimal
coding at the receiver with strong interference.

Finally, a note on separability. In \cite{cap_theorems:CadamJafar_Insep},
Cadambe and Jafar demonstrate the inseparability of parallel interference
channels using an example of a three-user frequency selective fading IFC. The
authors use interference alignment schemes to show that separability is not
optimal for fading IFCs with three or more users while leaving open the
question for the two-user fading IFC. We addressed this question in
\cite{cap_theorems:SXEP} for the ergodic fading one-sided IFC and developed
the conditions for the optimality of separability for EVS\ and US one-sided
IFCs. In this paper, we readdress this question for all sub-classes of fading
IFCs. Our results suggest that in general both one-sided and two-sided IFCs
benefit from transmitting the same information across all sub-channels, i.e.,
not independently encoding and decoding in each sub-channel, thereby
exploiting the fading diversity to mitigate interference.

The paper is organized as follows. In\ Section \ref{Section 2}, we present the
channel models studied. In Section \ref{section 3}, we summarize our main
results. The capacity region of an ergodic fading C-MAC is developed in
Section \ref{Sec_CM}. The proofs are collected in Section \ref{Sec_4}. We
discuss our results with numerical examples in Section \ref{Sec_Dis} and
conclude in Section \ref{Sec_Con}.%

\section{\label{Section 2}Channel Model and Preliminaries}%

\begin{figure}
[ptb]
\begin{center}
\includegraphics[
height=2.1655in,
width=5.9931in
]%
{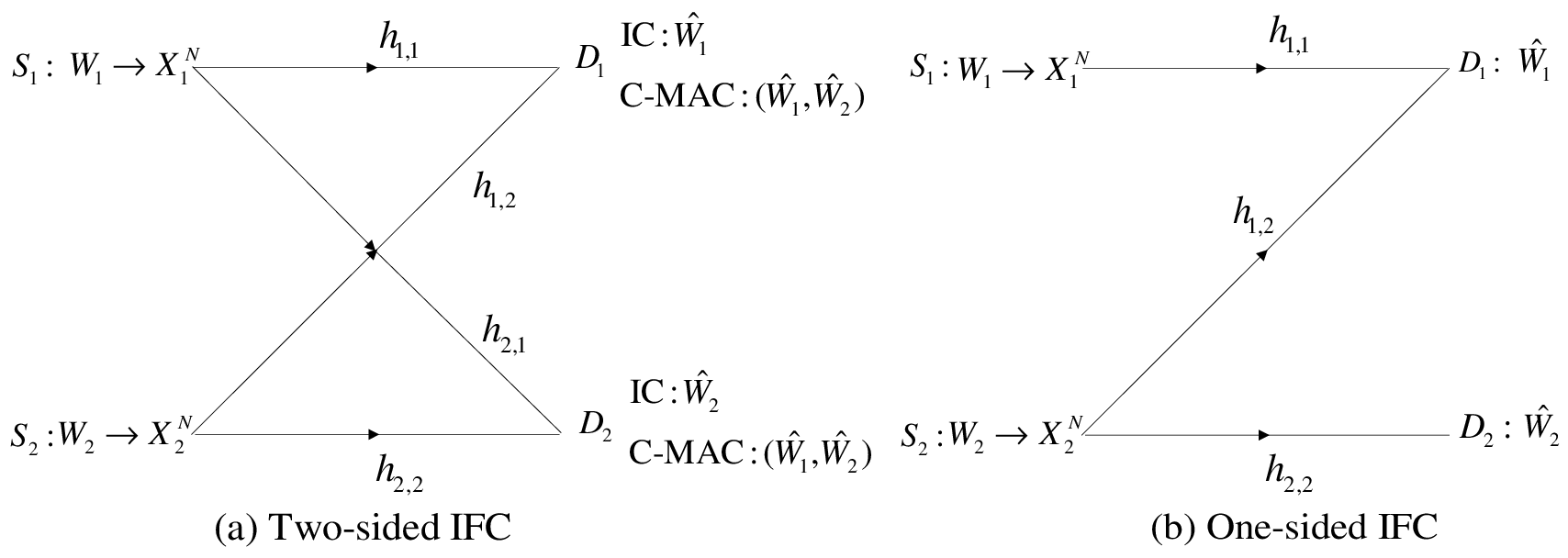}%
\caption{The two-user Gaussian\ two-sided IFC and C-MAC and the two-user
Gaussian one-sided IFC.}%
\label{Fig_IC}%
\end{center}
\end{figure}

A two-sender two-receiver (also referred to as the two-user) ergodic fading
Gaussian IFC consists of two source nodes $S_{1}$ and $S_{2}$, and two
destination nodes $D_{1}$ and $D_{2}$ as shown in Fig. \ref{Fig_IC}. Source
$S_{k}$, $k=1,2$, uses the channel $n$ times to transmit its message $W_{k}$,
which is distributed uniformly in the set $\,\{1,2,\ldots,2^{B_{k}}\}$ and is
independent of the message from the other source, to its intended receiver,
$D_{k}$, at a rate $R_{k}=B_{k}/n$ bits per channel use. In each use of the
channel, $S_{k}$ transmits the signal $X_{k}$ while the destination $D_{k}$
receives $Y_{k}$, $k=1,2.$ For $\mathbf{X}=\left[  X_{1}\text{ }X_{2}\right]
^{T}$, the channel output vector $\mathbf{Y}=\left[  Y_{1}\text{ }%
Y_{2}\right]  ^{T}$ is given by%
\begin{equation}
\mathbf{Y}=\mathbf{HX}+\mathbf{Z} \label{IC_Y}%
\end{equation}
where $\mathbf{Z}=\left[  Z_{1}\text{ }Z_{2}\right]  ^{T}$ is a noise vector
with entries that are zero-mean, unit variance, circularly symmetric complex
Gaussian noise variables and $\mathbf{H}$ is a random matrix of fading gains
with entries $H_{m,k}$, for all $m,k=1,2$, such that $H_{m,k}$ denotes the
fading gain between receiver $m$ and transmitter $k$. We use $\mathbf{h}$ to
denote a realization of $\mathbf{H}$. We assume the fading process $\left\{
\mathbf{H}\right\}  $ is stationary and ergodic but not necessarily Gaussian.
Note that the channel gains $H_{m,k}$, for all $m$ and $k$, are not assumed to
be independent; however, $\mathbf{H}$ is known instantaneously at all the
transmitters and receivers.

Over $n$ uses of the channel, the transmit sequences $\left\{  X_{k,i}%
\right\}  $ are constrained in power according to%
\begin{equation}
\left.  \sum\limits_{i=1}^{n}\left\vert X_{k,i}\right\vert ^{2}\leq
n\overline{P}_{k}\right.  ,\text{ for all }k=1,2\text{.} \label{IFC_Pwr}%
\end{equation}
Since the transmitters know the fading states of the links on which they
transmit, they can allocate their transmitted signal power according to the
channel state information. A power policy \underline{$P$}$(\mathbf{h})$ is a
mapping from the fading state space consisting of the set of all fading states
(instantiations) $\mathbf{h}$ to the set of non-negative real values in
$\mathcal{R}_{+}^{2}$. The entries of \underline{$P$}$(\mathbf{h})$ are
$P_{k}(\mathbf{h})$, the power policy at user $k$, $k=1,2$. While
\underline{$P$}$(\mathbf{h})$ denotes the map for a particular fading state,
we write \underline{$P$}$(\mathbf{H})$ to explicitly describe the policy for
the entire set of random fading states. Thus, we use the notation
\underline{$P$}$(\mathbf{H})$ when averaging over all fading states or
describing a collection of policies, one for every $\mathbf{h}$. The entries
of \underline{$P$}$(\mathbf{H})$ are $P_{k}(\mathbf{H}),$ for all $k$.

For an ergodic fading channel, (\ref{IFC_Pwr}) then simplifies to
\begin{equation}
\left.  \mathbb{E}\left[  P_{k}(\mathbf{H})\right]  \leq\overline{P}%
_{k}\right.  \text{ for all }k=1,2, \label{ErgPwr}%
\end{equation}
where the expectation in (\ref{ErgPwr}) is over the distribution of
$\mathbf{H}$. We denote the set of all feasible policies $\underline{P}\left(
\mathbf{h}\right)  $, i.e., the power policies whose entries satisfy
(\ref{ErgPwr}), by $\mathcal{P}$. Finally, we write \underline{$\overline{P}$}
to denote the vector of average power constraints with entries $\overline
{P}_{k}$, $k=1,2$.

For the special case where both receivers decode the messages from both
transmitters, we obtain a compound MAC (see Fig. \ref{Fig_IC}(a)). A one-sided
fading Gaussian IFC results when either $H_{1,2}=0$ or $H_{2,1}=0$ (see Fig.
\ref{Fig_IC}(b)). Without loss of generality, we develop sum-capacity results
for a one-sided IFC (Z-IFC) with $H_{2,1}=0$. The results extend naturally to
the complementary one-sided model with $H_{1,2}=0$. A two-sided IFC can be
viewed as a collection of two complementary one-sided IFCs, one with
$H_{1,2}=0$ and the other with $H_{2,1}=0$.

We write $\mathcal{C}_{\text{IFC}}\left(  \overline{P}_{1},\overline{P}%
_{2}\right)  $ and $\mathcal{C}_{\text{C-MAC}}\left(  \overline{P}%
_{1},\overline{P}_{2}\right)  $ to denote the capacity region of an ergodic
fading IFC and C-MAC, respectively. Our definition of average error
probabilities, capacity regions, and achievable rate pairs $\left(
R_{1},R_{2}\right)  $ for both the IFC and C-MAC mirror the standard
information-theoretic definitions \cite[Chap. 14]{cap_theorems:CTbook}.

Non-fading IFCs can be classified by the relative strengths of the interfering
to intended signals at each of the receivers. A (two-sided non-fading)
\textit{strong }IFC is one in which the cross-link channel gains are larger
than the direct link channel gains to the intended receivers
\cite{cap_theorems:Sato_IC}, i.e.,
\begin{equation}%
\begin{array}
[c]{cc}%
\left\vert H_{j,k}\right\vert >\left\vert H_{k,k}\right\vert  & \text{for all
}j,k=1,2,\text{ }j\not =k.
\end{array}
\label{IFC_Str}%
\end{equation}
A strong IFC is \textit{very strong} if the cross-link channel gains dominate
the transmit powers such that (see for e.g.,
\cite{cap_theorems:Sato_IC,cap_theorems:Carleial_VSIFC})%
\begin{equation}%
\begin{array}
[c]{cc}%
\sum\limits_{k=1}^{2}C\left(  \left\vert H_{k,k}\right\vert ^{2}P_{k}\left(
\mathbf{H}\right)  \right)  <C\left(  \sum\limits_{k=1}^{2}\left\vert
H_{j,k}\right\vert ^{2}P_{j}\left(  \mathbf{H}\right)  \right)  & \text{for
all }j=1,2,
\end{array}
\label{IFC_VStr}%
\end{equation}
where for the non-fading IFC, $P_{k}\left(  \mathbf{H}\right)  =\overline
{P}_{k}$ in (\ref{IFC_Pwr}). One can verify that (\ref{IFC_VStr}) implies
(\ref{IFC_Str}), i.e., a very strong IFC is also strong.

A non-fading IFC is \textit{weak} when (\ref{IFC_Str}) is not satisfied for
all $j,k$, i.e., neither of the two complementary one-sided IFCs that a
two-sided IFC can be decomposed into are strong. A non-fading IFC is
\textit{mixed} when one of complementary\ one-sided IFCs is weak while the
other is strong, i.e.,
\begin{equation}%
\begin{array}
[c]{ccc}%
\left\vert H_{1,2}\right\vert >\left\vert H_{2,2}\right\vert \text{ } &
\text{and } & \left\vert H_{2,1}\right\vert <\left\vert H_{1,1}\right\vert
\end{array}
\end{equation}
or
\begin{equation}%
\begin{array}
[c]{ccc}%
\left\vert H_{1,2}\right\vert >\left\vert H_{2,2}\right\vert \text{ } &
\text{and } & \left\vert H_{2,1}\right\vert <\left\vert H_{1,1}\right\vert .
\end{array}
\end{equation}

An ergodic fading IFC is a collection of parallel sub-channels (fading
states), and thus, each sub-channel can be either very strong, strong, or
weak. Since a fading IFC can contain a mixture of different types of
sub-channels, we introduce the following definitions to classify the set of
all ergodic fading two-user Gaussian IFCs (see also Fig. \ref{FigIFCVenn}).
Unless otherwise stated, we henceforth simply write IFC to denote a two-user
ergodic fading Gaussian IFC.

\begin{definition}
A \textit{uniformly strong} IFC is a collection of strong sub-channels, i.e.,
both cross-links in each sub-channel satisfy (\ref{IFC_Str}).
\end{definition}

\begin{definition}
An \textit{ergodic very strong} IFC is a collection of weak and strong
(including very strong) sub-channels for which (\ref{IFC_VStr}) is satisfied
when averaged over all fading states and for $P_{k}\left(  \mathbf{H}\right)
=P_{k}^{\left(  wf\right)  }\left(  H_{kk}\right)  $, where $P_{k}^{\left(
wf\right)  }\left(  H_{kk}\right)  $ is the optimal waterfilling policy that
achieves the point-to-point capacity for user $k$ in the absence of interference.
\end{definition}

\begin{definition}
A \textit{uniformly weak }IFC is a collection of weak sub-channels, i.e., in
each sub-channel both cross-links do not satisfy (\ref{IFC_Str}).
\end{definition}

\begin{definition}
A \textit{uniformly mixed }IFC is a pair of two complementary one-sided IFCs
in which one of them is uniformly weak and the other is uniformly strong.
\end{definition}

\begin{definition}
A \textit{hybrid }IFC is a collection of weak and strong sub-channels with at
least one weak and one strong sub-channel that do not satisfy the conditions
in (\ref{IFC_VStr}) when averaged over all fading states and for $P_{k}\left(
\mathbf{H}\right)  =P_{k}^{\left(  wf\right)  }\left(  H_{kk}\right)  $.
\end{definition}

Since an ergodic fading channel is a collection of parallel sub-channels
(fading states) with different weights, throughout the sequel, we use the
terms fading states and sub-channels interchangeably. In contrast to the
one-sided IFC, we simply write IFC to denote the two-sided model. Before
proceeding, we summarize the notation used in the sequel.

\begin{itemize}
\item Random variables (e.g. $H_{k,j}$) are denoted with uppercase letters and
their realizations (e.g. $h_{k,j}$) with the corresponding lowercase letters.

\item Bold font $\mathbf{X}$ denotes a random matrix while bold font
$\mathbf{x}$ denotes an instantiation of $\mathbf{X}$.

\item $I$ denotes the identity matrix.

\item $\left\vert \mathbf{X}\right\vert $ and $\mathbf{X}^{-1}$ denotes the
determinant and inverse of the matrix $\mathbf{X.}$

\item $\mathcal{CN}\left(  0,\mathbf{\Sigma}\right)  $ denotes a circularly
symmetric complex Gaussian distribution with zero mean and covariance
$\mathbf{\Sigma}$.

\item $\mathcal{K}=\left\{  1,2\right\}  $ denotes the set of transmitters.

\item $\mathbb{E}\left(  \cdot\right)  $ denotes expectation; $C(x)$ denotes
$\log(1+x)$ where the logarithm is to the base 2, $\left(  x\right)  ^{+}$
denotes $\max(x,0)$, $I(\cdot;\cdot)$ denotes mutual information, $h\left(
\cdot\right)  $ denotes differential entropy, and $R_{\mathcal{S}}$ denotes $%
{\textstyle\sum\nolimits_{k\in\mathcal{S}}}
R_{k}$ for any ${\mathcal{S}}$ $\subseteq\mathcal{K}$.
\end{itemize}

\section{\label{section 3}Main\ Results}

The following theorems summarize the main contributions of this paper. The
proof for the capacity region of the C-MAC is presented in Section
\ref{Sec_CM} as are the details of determining the capacity achieving power
policies. The proofs for the remaining theorems, related to IFCs, are
collected in\ Section \ref{Sec_4}. Throughout the sequel we write waterfilling
solution to denote the capacity achieving power policy for ergodic fading
point-to-point channels \cite{cap_theorems:GoldsmithVaraiya}.

\subsection{\label{Sec_3CM}Ergodic fading C-MAC}

An achievable rate region for ergodic fading IFCs results from allowing both
receivers to decode the messages from both transmitters, i.e., by converting
an IFC to a C-MAC. The following theorem summarizes the sum-capacity
$\mathcal{C}_{\text{C-MAC}}$ of an ergodic fading C-MAC.

\begin{theorem}
\label{Th_CMAC}The capacity region, $\mathcal{C}_{\text{C-MAC}}\left(
\overline{P}_{1},\overline{P}_{2}\right)  $, of an ergodic fading two-user
Gaussian\ C-MAC with average power constraints $P_{k}$ at transmitter $k$,
$k=1,2,$ is%
\begin{equation}
\mathcal{C}_{\text{C-MAC}}\left(  \overline{P}_{1},\overline{P}_{2}\right)
=\bigcup_{\underline{P}\in\mathcal{P}}\left\{  \mathcal{C}_{1}\left(
\underline{P}\left(  \mathbf{H}\right)  \right)  \cap\mathcal{C}_{2}\left(
\underline{P}\left(  \mathbf{H}\right)  \right)  \right\}  \label{CapR_CMAC}%
\end{equation}
where for $j=1,2,$ we have%
\begin{equation}
\mathcal{C}_{j}\left(  \underline{P}\left(  \mathbf{H}\right)  \right)
=\left\{  \left(  R_{1},R_{2}\right)  :R_{\mathcal{S}}\leq\mathbb{E}\left[
C\left(  \sum_{k\in\mathcal{S}}\left\vert H_{j,k}\right\vert ^{2}P_{k}\left(
\mathbf{H}\right)  \right)  \right]  ,\text{for all }\mathcal{S}%
\subseteq\mathcal{K}\right\}  . \label{CMAC_Cj}%
\end{equation}
The optimal coding scheme requires encoding and decoding jointly across all sub-channels.
\end{theorem}

\begin{remark}
The capacity region $\mathcal{C}_{\text{C-MAC}}$ is convex. This follows from
the convexity of the set $\mathcal{P}$ and the concavity of the $\log$ function.
\end{remark}

\begin{remark}
$\mathcal{C}_{\text{C-MAC}}$ is a function of $\left(  \overline{P}%
_{1},\overline{P}_{2}\right)  $ due to the fact that union in (\ref{CapR_CMAC}%
) is over all feasible power policies, i.e., over all $\underline{P}\left(
\mathbf{H}\right)  $ whose entries satisfy (\ref{ErgPwr}).
\end{remark}

\begin{remark}
In contrast to the ergodic fading point-to-point and multiple access channels,
the ergodic fading C-MAC is not merely a collection of independent parallel
channels; in fact encoding and decoding independently in each parallel channel
is in general sub-optimal as demonstrated later in the sequel.
\end{remark}

\begin{corollary}
\label{Cor_1}The capacity region $\mathcal{C}_{\text{IFC}}$ of an ergodic
fading IFC is bounded as $\mathcal{C}_{\text{C-MAC}}\subseteq\mathcal{C}%
_{\text{IFC}}$.
\end{corollary}

\subsection{Ergodic Very Strong IFCs}

\begin{theorem}
\label{Th_VS}The capacity region of an ergodic very strong IFC is
\begin{equation}
\mathcal{C}_{\text{IFC}}^{EVS}=\left\{  \left(  R_{1},R_{2}\right)  :R_{k}%
\leq\mathbb{E}\left[  C\left(  \left\vert H_{k,k}\right\vert ^{2}P_{k}%
^{wf}\left(  H_{k,k}\right)  \right)  \right]  ,k=1,2\right\}  .
\label{EVS_CapR}%
\end{equation}
The sum-capacity is
\begin{equation}
\sum_{k=1}^{2}\mathbb{E}\left[  C\left(  \left\vert H_{k,k}\right\vert
^{2}P_{k}^{wf}\left(  H_{k,k}\right)  \right)  \right]  \label{EVS_SC}%
\end{equation}
where, for all $k,$ $P_{k}^{wf}\left(  H_{j,k}\right)  $ is the optimal
waterfilling solution for an (interference-free) ergodic fading link between
transmitter $k$ and receiver $k$ such that, $\underline{P}^{wf}\left(
H_{k,k}\right)  $ satisfies
\begin{equation}
\sum_{k=1}^{2}\mathbb{E}\left[  C\left(  \left\vert H_{k,k}\right\vert
^{2}P_{k}^{wf}\left(  H_{k,k}\right)  \right)  \right]  <\min_{j=1,2}%
\mathbb{E}\left[  C\left(  \sum_{k=1}^{2}\left\vert H_{j,k}\right\vert
^{2}P_{k}^{wf}\left(  H_{k,k}\right)  \right)  \right]  . \label{EVS_Cond}%
\end{equation}
The capacity achieving scheme requires encoding and decoding jointly across
all sub-channels at the transmitters and receivers respectively. The optimal
strategy also requires both receivers to decode messages from both transmitters.
\end{theorem}

\begin{remark}
In the sequel we show that the condition in (\ref{EVS_Cond}) is a result of
the achievable strategy, and therefore is a sufficient condition. For the
special case of fixed (non-fading) channel gains $\mathbf{H}$, and
$P_{k}^{\ast}=\overline{P}_{1}$, (\ref{EVS_Cond}) reduces to the general
conditions for a very strong IFC (see for e.g., \cite{cap_theorems:Sato_IC})
given by
\begin{subequations}
\label{VS_NF_IFC}%
\begin{align}
\left\vert H_{1,2}\right\vert ^{2}  &  >\left\vert H_{2,2}\right\vert
^{2}\left(  1+\left\vert H_{1,1}\right\vert ^{2}\overline{P}_{1}\right) \\
\left\vert H_{2,1}\right\vert ^{2}  &  >\left\vert H_{1,1}\right\vert
^{2}\left(  1+\left\vert H_{2,2}\right\vert ^{2}\overline{P}_{2}\right)  .
\end{align}
In contrast, the fading averaged conditions in (\ref{EVS_Cond}) imply that not
every sub-channel needs to satisfy (\ref{VS_NF_IFC}) and in fact, the ergodic
very strong channel can be a mix of weak and strong channels provided
$\underline{P}^{\left(  wf\right)  }$ satisfies (\ref{EVS_Cond}). This in turn
implies that not every parallel sub-channel needs to be a strong
(non-fading)\ Gaussian IFC.
\end{subequations}
\end{remark}

\begin{remark}
The set of strong fading IFCs for which every sub-channel is strong and the
optimal waterfilling policies for the two interference-free links satisfy
(\ref{EVS_Cond}) is strictly a subset of the set of ergodic very strong IFCs.
\end{remark}

\begin{remark}
As stated in Theorem \ref{Th_VS}, the capacity achieving scheme for EVS\ IFCs
requires coding jointly across all sub-channels. Coding independent messages
(separable coding) across the sub-channels is optimal only when every
sub-channel is very strong at the optimal policy $\underline{P}^{\left(
wf\right)  }$.
\end{remark}

\subsection{Uniformly Strong IFC}

In the following theorem, we present the capacity region and the sum-capacity
of a uniformly strong IFC.

\begin{theorem}
\label{Th_Str}The capacity region of a uniformly strong fading IFC for which
the entries of every fading state $\mathbf{h}$ satisfy%
\begin{equation}%
\begin{array}
[c]{ccc}%
\left\vert h_{1,1}\right\vert \leq\left\vert h_{2,1}\right\vert  & \text{and}
& \left\vert h_{2,2}\right\vert \leq\left\vert h_{1,2}\right\vert
\end{array}
\label{US_HCond}%
\end{equation}
is given by
\begin{equation}
\mathcal{C}_{\text{IFC}}^{US}\left(  \overline{P}_{1},\overline{P}_{2}\right)
=\mathcal{C}_{\text{C-MAC}}\left(  \overline{P}_{1},\overline{P}_{2}\right)
\end{equation}
where $\mathcal{C}_{\text{C-MAC}}\left(  \overline{P}_{1},\overline{P}%
_{2}\right)  $ is the capacity of an ergodic fading C-MAC with the same
channel statistics as the IFC. The sum-capacity is
\begin{equation}
\max_{\underline{P}\left(  \mathbf{H}\right)  \in\mathcal{P}}\min\left\{
\min_{j=1,2}\left\{  \mathbb{E}\left[  C\left(
{\textstyle\sum\nolimits_{k=1}^{2}}
\left\vert H_{j,k}\right\vert ^{2}P_{k}\left(  \mathbf{H}\right)  \right)
\right]  \right\}  ,\sum_{k=1}^{2}\mathbb{E}\left[  C\left(  \left\vert
H_{k,k}\right\vert ^{2}P_{k}\left(  \mathbf{H}\right)  \right)  \right]
\right\}  . \label{US_SC}%
\end{equation}
The capacity achieving scheme requires encoding and decoding jointly across
all sub-channels at the transmitters and receivers, respectively, and also
requires both receivers to decode messages from both transmitters.
\end{theorem}

\begin{remark}
In contrast to the very strong case, every sub-channel in a uniformly strong
fading IFC is strong.
\end{remark}

\begin{remark}
\label{Rem_USSep}The uniformly strong condition may suggest that separability
is optimal. However, the capacity achieving C-MAC approach requires joint
encoding and decoding across all sub-channels. A strategy where each
sub-channel is viewed as an independent IFC, as in
\cite{cap_theorems:ChuCioffi_IC}, will in general be strictly sub-optimal.
This is seen directly from comparing (\ref{US_SC}) with the sum-rate achieved
by coding independently over the sub-channels which is given by%
\begin{equation}
\max_{\underline{P}\left(  \mathbf{H}\right)  \in\mathcal{P}}\mathbb{E}%
\left\{  \min\left\{  \min_{j=1,2}\left\{  C\left(
{\textstyle\sum\nolimits_{k=1}^{2}}
\left\vert H_{j,k}\right\vert ^{2}P_{k}\left(  \mathbf{H}\right)  \right)
\right\}  \right.  \right.  ,\left.  \left.  \sum_{k=1}^{2}C\left(  \left\vert
H_{k,k}\right\vert ^{2}P_{k}\left(  \mathbf{H}\right)  \right)  \right\}
\right\}  .\text{ \ \ \ \ \ \ \ \ \ \ } \label{US_Ach}%
\end{equation}

\end{remark}

The sub-optimality of independent encoding follows directly from the fact that
for two random variables $A\left(  \mathbf{H}\right)  $ and $\not B \left(
\mathbf{H}\right)  ,$ $\mathbb{E}[\min\left(  A\left(  \mathbf{H}\right)
,B\left(  \mathbf{H}\right)  \right)  ]$ $\leq$ $\min\left(  \mathbb{E}%
[A\left(  \mathbf{H}\right)  ],\mathbb{E}[B\left(  \mathbf{H}\right)
]\right)  ]$ with equality \textit{if and only if} for every fading
instantiation $\mathbf{h}$, $A\left(  \mathbf{H}\right)  $ (resp. $B\left(
\mathbf{H}\right)  $) dominates $B\left(  \mathbf{H}\right)  $ (resp.
$A\left(  \mathbf{H}\right)  $). Thus, independent (separable) encoding across
sub-channels is optimal only when, at $\underline{P}^{\ast}\left(
\mathbf{H}\right)  $, the sum-rate in every sub-channel in (\ref{US_Ach}) is
maximized by the same sum-rate function.

\subsection{Uniformly Weak One-Sided IFC}

The following theorem summarizes the sum-capacity of a one-sided uniformly
weak IFC in which every sub-channel is weak.

\begin{theorem}
\label{Th_UW1}The sum-capacity of a uniformly weak ergodic fading Gaussian
one-sided IFC for which the entries of every fading state $\underline{h}$
satisfy%
\begin{equation}%
\begin{array}
[c]{c}%
\left\vert h_{2,2}\right\vert >\left\vert h_{1,2}\right\vert
\end{array}
\label{UW_Cond}%
\end{equation}
is given by
\begin{equation}
\max_{\underline{P}\left(  \mathbf{H}\right)  \in\mathcal{P}}\left\{
S^{\left(  w,1\right)  }\left(  \underline{P}\left(  \mathbf{H}\right)
\right)  \right\}  \label{SC_Weak}%
\end{equation}
where
\begin{equation}
S^{\left(  w,1\right)  }\left(  \underline{P}\left(  \mathbf{H}\right)
\right)  =\mathbb{E}\left[  C\left(  \frac{\left\vert H_{1,1}\right\vert
^{2}P_{1}\left(  \mathbf{H}\right)  }{1+\left\vert H_{1,2}\right\vert
^{2}P_{2}\left(  \mathbf{H}\right)  }\right)  +C\left(  \left\vert
H_{2,2}\right\vert ^{2}P_{2}\left(  \mathbf{H}\right)  \right)  \right]  .
\label{SCW_S}%
\end{equation}

\end{theorem}

\begin{remark}
One could alternately consider the fading one-sided IFC in which $\left\vert
h_{1,1}\right\vert >\left\vert h_{2,1}\right\vert $ and $h_{1,2}=0$ for the
sum-capacity is given by (\ref{SC_Weak}) with the superscript $1$ replaced by
2. The expression $S^{\left(  w,2\right)  }\left(  \underline{P}\left(
\mathbf{H}\right)  \right)  $ is given by (\ref{SCW_S}) after swapping the
indexes $1$ and $2$.
\end{remark}

\subsection{Uniformly Mixed IFC}

The following theorem summarizes the sum-capacity of a class of uniformly
mixed two-sided IFC.

\begin{theorem}
\label{Th_Mix}For a class of uniformly mixed ergodic fading two-sided Gaussian
IFCs for which the entries of every fading state $\underline{h}$ satisfy%
\begin{equation}%
\begin{array}
[c]{ccc}%
\left\vert h_{1,1}\right\vert >\left\vert h_{2,1}\right\vert  & \text{and} &
\left\vert h_{2,2}\right\vert \leq\left\vert h_{1,2}\right\vert
\end{array}
\end{equation}
the sum-capacity is
\begin{equation}
\max_{\underline{P}\left(  \mathbf{H}\right)  \in\mathcal{P}}\left\{
\min\left(  \mathbb{E}\left[  C\left(
{\textstyle\sum\nolimits_{k=1}^{2}}
\left\vert H_{1,k}\right\vert ^{2}P_{k}\left(  \mathbf{H}\right)  \right)
\right]  ,S^{\left(  w,2\right)  }\left(  \underline{P}\left(  \mathbf{H}%
\right)  \right)  \right)  \right\}  \label{SC_Mix}%
\end{equation}
where $S^{\left(  w,2\right)  }\left(  \underline{P}\left(  \mathbf{H}\right)
\right)  $ is given by (\ref{SCW_S}) by swapping indexes $1$ and $2$.
\end{theorem}

\begin{remark}
One could alternately consider the fading IFC in which $\left\vert
h_{1,1}\right\vert \leq\left\vert h_{2,1}\right\vert $ and $\left\vert
h_{2,2}\right\vert >\left\vert h_{1,2}\right\vert $. The sum-capacity is given
by (\ref{SC_Mix}) after swapping the indexes $1$ and $2$.
\end{remark}

\begin{remark}
For the special case of $H_{k,k}=\sqrt{SNR}e^{j\phi_{kk}}$ and $H_{j,k}%
=\sqrt{INR}e^{j\phi_{jk}}$, $j\not =k$, where $\phi_{j,k}$ for all $j$ and $k$
is independent and distributed uniformly in $\left[  -\pi,\pi\right]  $, the
sum-capacity in Theorems \ref{Th_Str} and \ref{Th_Mix} can also be achieved by
ergodic interference alignment as shown in \cite{cap_theorems:Jafar_ErgIFC}.
\end{remark}

\subsection{Uniformly Weak IFC}

The sum-capacity of a one-sided uniformly weak IFC in Theorem \ref{Th_UW1} is
an upper bound for that of a two-sided IFC for which at least one of two
one-sided IFCs that result from eliminating a cross-link is uniformly weak.
Similarly, a bound can be obtained from the sum-capacity of the complementary
one-sided IFC. The following theorem summarizes this result.

\begin{theorem}
\label{Th_UW2}For a class of uniformly weak ergodic fading two-sided Gaussian
IFCs for which the entries of every fading state $\underline{h}$ satisfy%
\begin{equation}%
\begin{array}
[c]{ccc}%
\left\vert h_{1,1}\right\vert >\left\vert h_{2,1}\right\vert  & \text{and} &
\left\vert h_{2,2}\right\vert >\left\vert h_{1,2}\right\vert
\end{array}
\end{equation}
the sum-capacity is upper bounded as
\begin{equation}
R_{1}+R_{2}\leq\max_{\underline{P}\left(  \mathbf{H}\right)  \in\mathcal{P}%
}\min\left(  S^{\left(  w,1\right)  }\left(  \underline{P}\left(
\mathbf{H}\right)  \right)  ,S^{\left(  w,2\right)  }\left(  \underline
{P}\left(  \mathbf{H}\right)  \right)  \right)  . \label{SC_UW2}%
\end{equation}

\end{theorem}

\begin{remark}
For the non-fading case, the sum-rate bounds in (\ref{SC_UW2}) simplify to
those obtained in \cite[Theorem 3]{cap_theorems:ETW}.
\end{remark}

\subsection{One-sided IFC: General Achievable Scheme}

For EVS and US IFCs, Theorems \ref{Th_VS} and \ref{Th_Str} suggest that joint
coding across all sub-channels is optimal. Particularly for EVS, such joint
coding allows one to exploit the strong states in decoding messages. Relying
on this observation, we present an achievable strategy based on joint coding
all sub-classes of one-sided IFCs with $H_{2,1}=0$. The encoding scheme
involves rate-splitting at user $2$, i.e., user $2$ transmits $w_{2}=\left(
w_{2p},w_{2c}\right)  $ where $w_{2p}$ and $w_{2c}$ are private and common
messages, respectively and can be viewed as a Han-Kobayashi scheme with
Gaussian codebooks and without time-sharing.

\begin{theorem}
\label{Th_Hyb}The sum-capacity of a one-sided IFC is lower bounded by%
\begin{equation}
\max_{\underline{P}\left(  \mathbf{H}\right)  \in\mathcal{P},\alpha
_{\mathbf{H}}\in\lbrack0,1]}\min\left(  S_{1}\left(  \alpha_{\mathbf{H}%
},\underline{P}\left(  \mathbf{H}\right)  \right)  ,S_{2}\left(
\alpha_{\mathbf{H}},\underline{P}\left(  \mathbf{H}\right)  \right)  \right)
\label{HK1_SR}%
\end{equation}
where
\begin{align}
S_{1}\left(  \alpha_{\mathbf{H}},\underline{P}\left(  \mathbf{H}\right)
\right)   &  =\mathbb{E}\left[  C\left(  \frac{\left\vert H_{1,1}\right\vert
^{2}P_{1}\left(  \mathbf{H}\right)  }{1+\left\vert H_{1,2}\right\vert
^{2}\alpha_{\mathbf{H}}P_{2}\left(  \mathbf{H}\right)  }\right)  \right]
+\mathbb{E}\left[  C\left(  \left\vert H_{2,2}\right\vert ^{2}P_{2}\left(
\mathbf{H}\right)  \right)  \right]  ,\\
S_{2}\left(  \alpha_{\mathbf{H}},\underline{P}\left(  \mathbf{H}\right)
\right)   &  =\mathbb{E}\left[  C\left(  \left\vert H_{2,2}\right\vert
^{2}\alpha_{\mathbf{H}}P_{2}\left(  \mathbf{H}\right)  \right)  \right]
+\mathbb{E}\left[  C\left(  \frac{\left\vert H_{1,1}\right\vert ^{2}%
P_{1}\left(  \mathbf{H}\right)  +\left\vert H_{1,2}\right\vert ^{2}%
\overline{\alpha}_{\mathbf{H}}P_{2}\left(  \mathbf{H}\right)  }{1+\left\vert
H_{1,2}\right\vert ^{2}\alpha_{\mathbf{H}}P_{2}\left(  \mathbf{H}\right)
}\right)  \right]  ,
\end{align}
such that $\alpha_{\mathbf{H}}$ is the power allocated by user $2$ in fading
state $\mathbf{H}$ to transmitting $w_{2p}$ and $\overline{\alpha}%
_{\mathbf{H}}=1-\alpha_{\mathbf{H}}$, $\alpha_{\mathbf{H}}\in\left[
0,1\right]  $. For EVS\ one-sided IFCs, the sum-capacity is achieved by
choosing $\alpha_{\mathbf{H}}=0$ for all $\mathbf{H}$ provided $S_{1}\left(
0,\underline{P}^{(wf)}\left(  \mathbf{H}\right)  \right)  <S_{2}\left(
0,\underline{P}^{(wf)}\left(  \mathbf{H}\right)  \right)  $. For US one-sided
IFCs, the sum-capacity is given by (\ref{HK1_SR}) for $\alpha_{\mathbf{H}}=0$
for all $\mathbf{H}$. For UW\ one-sided IFCs, the sum-capacity is achieved by
choosing $\alpha_{\mathbf{H}}=1$ and maximizing $S_{2}\left(  1,\underline
{P}\left(  \mathbf{H}\right)  \right)  =S_{1}\left(  1,\underline{P}\left(
\mathbf{H}\right)  \right)  $ over all feasible $\underline{P}\left(
\mathbf{H}\right)  .$ For a hybrid one-sided IFC, the achievable sum-rate is
maximized by
\begin{equation}
\alpha_{\mathbf{H}}^{\ast}=\left\{
\begin{array}
[c]{cc}%
\alpha\left(  \mathbf{H}\right)  \in(0,1] & \text{sub-channel }\mathbf{H}%
\text{ is weak }\\
0 & \text{sub-channel }\mathbf{H}\text{ is strong.}%
\end{array}
\right.  \label{alpstar_hyb}%
\end{equation}
and is given by (\ref{HK1_SR}) for this choice of $\alpha_{\mathbf{H}}^{\ast}$.
\end{theorem}

\begin{remark}
The optimal $\alpha_{\mathbf{H}}^{\ast}$ in (\ref{alpstar_hyb}) implies that
in general for the hybrid one-sided IFCs joint coding the transmitted message
across all sub-channels is optimal. Specifically, the common message is
transmitted jointly in all sub-channels while the private message is
transmitted only in the weak sub-channels. 
\end{remark}

\begin{remark}
The separation-based coding scheme of \cite{cap_theorems:SumCap_Par_ZIFC} is a
special case of the above HK-based coding scheme and is obtained by choosing
$\alpha_{\mathbf{H}}=1$ and $\alpha_{\mathbf{H}}=0$ for the weak and strong
states, respectively. The resulting sum-rate is at most as large as the bound
in (\ref{HK1_SR}) obtained for $\alpha_{\mathbf{H}}^{\ast}\in(0,1]$ and
$\alpha_{\mathbf{H}}^{\ast}=0$ for the weak and strong states, respectively.
\end{remark}

\begin{remark}
In \cite{cap_theorems:Tuninetti}, a Han-Kobayashi based scheme using Gaussian
codebooks and no time-sharing is used to develop an inner bound on the
capacity region of a two-sided IFC.
\end{remark}

\section{\label{Sec_CM}Compound MAC: Capacity Region and Optimal Policies}

As stated in\ Corollary \ref{Cor_1}, an inner bound on the sum-capacity of an
IFC can be obtained by allowing both receivers to decode both messages, i.e.,
by determining the sum-capacity of a C-MAC with the same inter-node links. In
this Section, we prove Theorem \ref{Th_CMAC} which establishes the capacity
region of ergodic fading C-MACs and discuss the optimal power policies that
achieve every point on the boundary of the capacity region.

\subsection{Capacity Region}

The capacity region of a discrete memoryless compound MAC is developed in
\cite{cap_theorems:Ahlswede_CMAC}. For each choice of input distribution at
the two independent sources, this capacity region is an intersection of the
MAC capacity regions achieved at the two receivers. The techniques in
\cite{cap_theorems:Ahlswede_CMAC} can be easily extended to develop the
capacity region for a Gaussian C-MAC with fixed channel gains. For the
Gaussian C-MAC, one can show that Gaussian signaling achieves the capacity
region using the fact that Gaussian signaling maximizes the MAC region at each
receiver. Thus, the Gaussian C-MAC capacity region is an intersection of the
Gaussian\ MAC capacity regions achieved at $D_{1}$ and $D_{2}$. For a
stationary and ergodic process $\left\{  \mathbf{H}\right\}  $, the channel in
(\ref{IC_Y}) can be modeled as a parallel Gaussian C-MACs consisting of a
collection of independent Gaussian C-MACs, one for each fading state
$\mathbf{h}$, with an average transmit power constraint over all parallel channels.

We now prove Theorem \ref{Th_CMAC} stated in\ Section \ref{Sec_3CM} which
gives the capacity region of ergodic fading C-MACs.

\textit{Proof of Theorem }\ref{Th_CMAC}

We first present an achievable scheme. Consider a policy $\underline{P}\left(
\mathbf{H}\right)  \in\mathcal{P}$. The achievable scheme involves requiring
each transmitter to encode the same message across all sub-channels and each
receiver to jointly decode over all sub-channels. Independent codebooks are
used for every sub-channel. An error occurs at receiver $j$ if one or both
messages decoded jointly across all sub-channels is different from the
transmitted message. Given this encoding and decoding, the analysis at each
receiver mirrors that for a MAC receiver \cite[14.3]{cap_theorems:CTbook} and
one can easily verify that for reliable reception of the transmitted message
at receiver $j$, the rate pair $\left(  R_{1},R_{2}\right)  $ needs to satisfy
the rate constraints in (\ref{CMAC_Cj}) where in decoding $w_{\mathcal{S}%
}=\left\{  w_{k}:k\in\mathcal{S}\right\}  $ the information collected in each
sub-channel is given by $C\left(  \sum_{k\in\mathcal{S}}\left\vert
H_{j,k}\right\vert ^{2}P_{k}\left(  \mathbf{H}\right)  \right)  $, for all
$\mathcal{S}\subseteq\mathcal{K}.$ Thus, for any feasible $\underline
{P}\left(  \mathbf{H}\right)  $, the achievable rate region is given by
$\mathcal{C}_{1}\left(  \underline{P}\left(  \mathbf{H}\right)  \right)
\cap\mathcal{C}_{2}\left(  \underline{P}\left(  \mathbf{H}\right)  \right)  $.
From the concavity of the $\log$ function, the achievable region over all
$\underline{P}\left(  \mathbf{H}\right)  $ is given by (\ref{CapR_CMAC}%
).\newline\qquad For the converse, the proof technique mirrors the proof for
the capacity of an ergodic fading MAC developed in \cite[Appendix
A]{cap_theorems:TH01}. For any $\underline{P}\left(  \mathbf{H}\right)
\in\mathcal{P}$, one can using similar limiting arguments to show that for
asymptotically error-free performance at receiver $j$, for all $j$, the
achievable region has to be bounded as%
\begin{equation}%
\begin{array}
[c]{cc}%
R_{\mathcal{S}}\leq\mathbb{E}\left[  C\left(  \sum_{k\in\mathcal{S}}\left\vert
H_{j,k}\right\vert ^{2}P_{k}\left(  \mathbf{H}\right)  \right)  \right]  , &
j=1,2.
\end{array}
\label{CMAC_OB}%
\end{equation}
The proof is completed by noting that due to the concavity of the $\log$ it
suffices to take the union of the region over all $\underline{P}\left(
\mathbf{H}\right)  \in\mathcal{P}$.

\begin{remark}
An achievable scheme in which independent messages are encoded in each
sub-channel, i.e., separable coding, will in general not achieve the capacity
region. This is due to the fact that for this separable coding scheme the
achievable rate in each sub-channel is a minimum of the rates at each
receiver. The average of such minima can at most be the minimum of the average
rates at each receiver, where the latter is achieved by encoding the same
message jointly across all sub-channels.
\end{remark}

Corollary \ref{Cor_1} follows from the argument that a rate pair in
$\mathcal{C}_{\text{C-MAC}}$ is achievable for the IFC since $\mathcal{C}%
_{\text{C-MAC}}$ is the capacity region when both messages are decoded at both receivers.%

\begin{figure*}[tbp] \centering
{\includegraphics[
height=2.0911in,
width=5.3056in
]%
{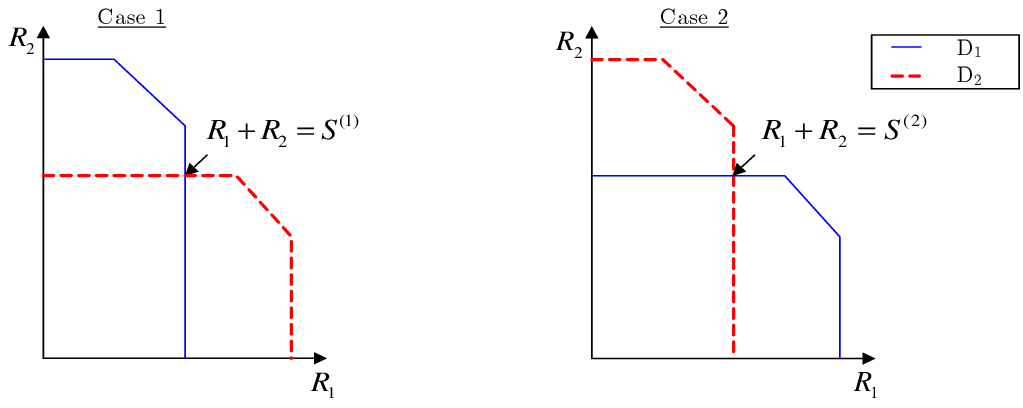}%
}%
\caption{Rate regions $\mathcal{C}_{1}(\underline{P}(\underline{H}%
))$ and $\mathcal{C}_{2}(\underline{P}(\underline{H}%
))$ and sum-rate for case 1 and case 2.}%
\label{Fig_Case12}%
\end{figure*}%
%

\begin{figure*}[tbp] \centering
{\includegraphics[
height=1.9268in,
width=5.3333in
]%
{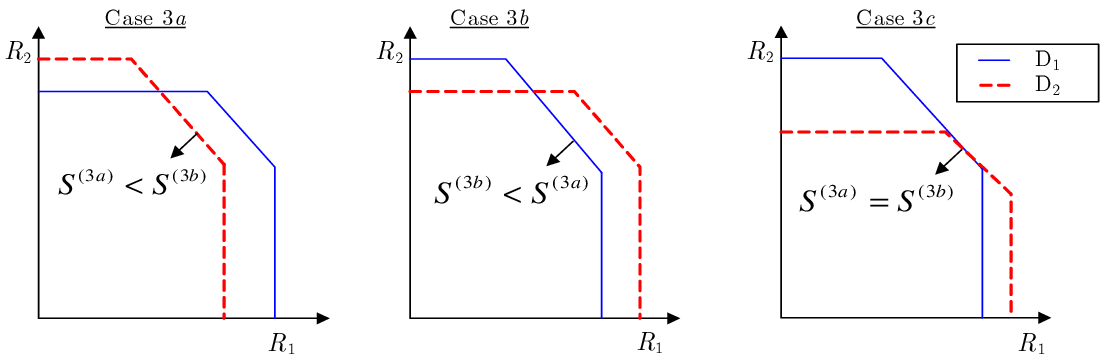}%
}%
\caption{Rate regions $R_{r}(\underline{P}(\underline{H}))$ and $R_{d}(\underline{P}(\underline{H}))$ and sum-rate for cases $3a$, $3b$, and $3c$.}\label{Fig_Case3abc}%
\end{figure*}%

\subsection{Sum-Capacity Optimal Policies}

The capacity region $\mathcal{C}_{\text{C-MAC}}$ is a union of the
intersection of the pentagons $\mathcal{C}_{1}\left(  \underline{P}\left(
\mathbf{H}\right)  \right)  $ and $\mathcal{C}_{2}\left(  \underline{P}\left(
\mathbf{H}\right)  \right)  $ achieved at $D_{1}$ and $D_{2},$ respectively,
where the union is over all $\underline{P}\left(  \mathbf{H}\right)
\in\mathcal{P}$. The region $\mathcal{C}_{\text{C-MAC}}$ is convex, and thus,
each point on the boundary of $\mathcal{C}_{\text{C-MAC}}$ is obtained by
maximizing the weighted sum $\mu_{1}R_{1}$ $+$ $\mu_{2}R_{2}$ over all
$\underline{P}\left(  \mathbf{H}\right)  \in\mathcal{P}$, and for all $\mu
_{1}>0$, $\mu_{2}>0$, subject to (\ref{CMAC_OB}). In this section, we
determine the optimal policy $\underline{P}^{\ast}\left(  \mathbf{H}\right)  $
that maximizes the sum-rate $R_{1}+R_{2}$ when $\mu_{1}$ $=$ $\mu_{2}$ $=$
$1$. Using the fact that the rate regions $\mathcal{C}_{1}\left(
\underline{P}\left(  \mathbf{H}\right)  \right)  $ and $\mathcal{C}_{2}\left(
\underline{P}\left(  \mathbf{H}\right)  \right)  $, for any feasible
$\underline{P}\left(  \mathbf{H}\right)  $, are pentagons, in Figs.
\ref{Fig_Case12} and \ref{Fig_Case3abc} we illustrate the five possible
choices for the sum-rate resulting from an intersection of $\mathcal{C}%
_{1}\left(  \underline{P}\left(  \mathbf{H}\right)  \right)  $ and
$\mathcal{C}_{2}\left(  \underline{P}\left(  \mathbf{H}\right)  \right)  $
(see also \cite{cap_theorems:SankarLiang_Conf}).

Cases $1$ and $2$, as shown in Fig. \ref{Fig_Case12} and henceforth referred
to as \textit{inactive cases}, are such that the constraints on the two
sum-rates are not active in $\mathcal{C}_{1}\left(  \underline{P}\left(
\mathbf{H}\right)  \right)  \cap\mathcal{C}_{2}\left(  \underline{P}\left(
\mathbf{H}\right)  \right)  $, i.e., no rate tuple on the sum-rate plane
achieved at one of the receivers lies within or on the boundary of the rate
region achieved at the other receiver. In contrast, when there exists at least
one such rate tuple such that the two sum-rates constraints are active in
$\mathcal{C}_{1}\left(  \underline{P}\left(  \mathbf{H}\right)  \right)
\cap\mathcal{C}_{2}\left(  \underline{P}\left(  \mathbf{H}\right)  \right)  $
are the \textit{active cases}. This includes Cases $3a$, $3b$, and $3c$ shown
in Fig. \ref{Fig_Case3abc} where the sum-rate at $D_{1}$ is smaller, larger,
or equal, respectively, to that achieved at $D_{2}$. By definition, the active
set also include the \textit{boundary cases} in which there is exactly one
rate pair that lies within or on the boundary of the rate region achieved at
the other receiver.\ There are six possible boundary cases that lie at the
intersection of an inactive case $l$, $l=1,2,$ and an active case $n,$
$n=3a,3b,3c$. There are six such boundary cases that we denote as cases
$\left(  l,n\right)  $, $l=1,2,$ and $n=3a,3b,3c$.

\begin{figure*}[tbp] \centering
{\includegraphics[
height=2.3462in,
width=5.8055in
]%
{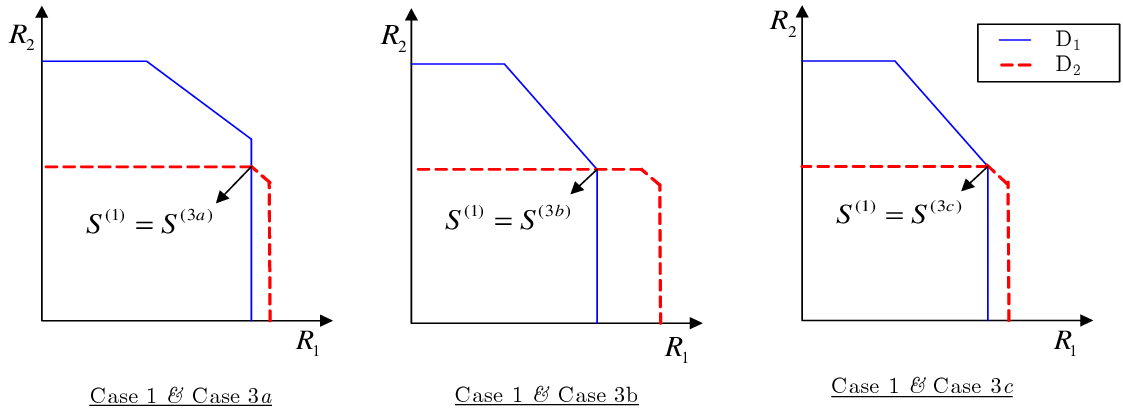}%
}%
\caption{Rate regions $R_{r}(\underline{P}(\underline{H}))$ and $R_{d}(\underline{P}(\underline{H}))$ for cases (1,3a), (1,3b), and (1,3c).}\label{Fig_BC13}%
\end{figure*}

\begin{figure*}[tbp] \centering
{\includegraphics[
height=2.3333in,
width=5.6247in
]%
{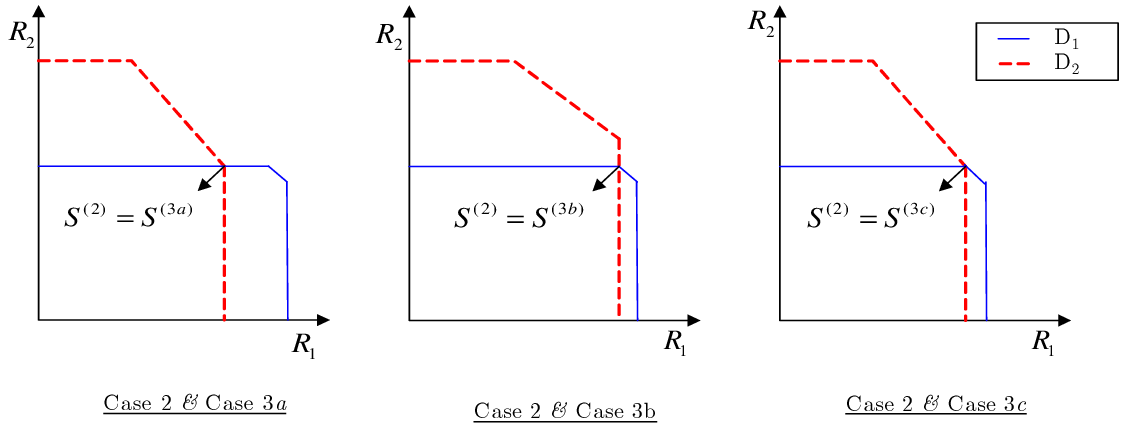}%
}%
\caption{Rate regions $R_{r}(\underline{P}(\underline{H}))$ and $R_{d}(\underline{P}(\underline{H}))$ for cases (2,3a), (2,3b), and (2,3c).}\label{Fig_BC23}%
\end{figure*}

In general, it is not possible to know \textit{a priori} the type of
intersection that will maximize the sum-capacity. Thus, the sum-rate for each
case has to be maximized over all $\underline{P}\left(  \mathbf{H}\right)
\in\mathcal{P}$. To simplify optimization and obtain a unique solution, we
explicitly consider the six boundary cases as distinct from the active cases
thereby ensuring that the subsets of power policies resulting in the different
cases are disjoint, i.e., no power policy results in more than one case. This
in turn implies that the power policies resulting in each case satisfy
specific conditions that distinguish that case from all others. For example,
from Fig. \ref{Fig_Case12}, Case 1 results only when $\sum\nolimits_{k=1}%
^{2}C\left(  H_{kk}P_{k}^{\left(  wf\right)  }\left(  \mathbf{H}\right)
\right)  <C\left(  \sum\nolimits_{k=1}^{2}H_{j,k}P_{k}^{\left(  wf\right)
}\left(  \mathbf{H}\right)  \right)  $, for all $j=1,2.$ Using these disjoint
cases and the fact that the rate expressions in (\ref{CMAC_OB}) are concave
functions of $\underline{P}\left(  \mathbf{H}\right)  $ allows us to develop
closed form sum-capacity results and optimal policies for all cases. Observe
that cases $1$ and $2$ do not share a boundary since such a transition (see
Fig. \ref{Fig_Case12}) requires passing through case $3a$ or $3b$ or $3c$.
Finally, note that Fig. \ref{Fig_Case3abc} illustrates two specific
$\mathcal{C}_{1}$ and $\mathcal{C}_{2}$ regions for $3a$, $3b$, and $3c$. The
conditions for each case are shown in Figs. \ref{Fig_Case12}-\ref{Fig_BC23}.%

%

Let $\underline{P}^{(i)}(\mathbf{H})$ and $\underline{P}^{(l,n)}(\mathbf{H})$
denote the optimal policies for cases $i$ and $(l,n)$, respectively. Let
$S^{\left(  i\right)  }(\underline{P}(\mathbf{H}))$ and $S^{\left(
l,n\right)  }(\underline{P}(\mathbf{H}))$ denote the sum-rate achieved for
cases $i$ and $\left(  l,n\right)  $, respectively, for some $\underline
{P}(\mathbf{H})\in\mathcal{P}$. The optimization problem for case $i$ or case
$\left(  l,n\right)  $ is given by
\begin{equation}%
\begin{array}
[c]{l}%
\max\limits_{\underline{P}\left(  \mathbf{H}\right)  \in\mathcal{P}}S^{\left(
i\right)  }\left(  \underline{P}\left(  \mathbf{H}\right)  \right)  \text{ or
}\max\limits_{\underline{P}\left(  \mathbf{H}\right)  \in\mathcal{P}%
}S^{\left(  l,n\right)  }\left(  \underline{P}\left(  \mathbf{H}\right)
\right) \\%
\begin{array}
[c]{ccc}%
s.t. & \mathbb{E}\left[  P_{k}(\mathbf{H})\right]  \leq\overline{P}%
_{k}\text{,} & \text{ }k=1,2,
\end{array}
\\%
\begin{array}
[c]{ccc}%
\text{ \ \ \ \ \ } & P_{k}(\mathbf{H})\geq0\text{,} & \text{ }k=1,2,\text{ for
all }\mathbf{H}%
\end{array}
\end{array}
\label{DF_OptProb}%
\end{equation}
where
\begin{equation}%
\begin{array}
[c]{l}%
\begin{array}
[c]{cc}%
S^{\left(  1\right)  }\left(  \underline{P}\left(  \mathbf{H}\right)  \right)
=\sum_{k=1}^{2}\mathbb{E}\left[  C\left(  \left\vert H_{k,k}\right\vert
^{2}P_{k}\left(  \mathbf{H}\right)  \right)  \right]  &
\end{array}
\\%
\begin{array}
[c]{cc}%
S^{\left(  2\right)  }\left(  \underline{P}\left(  \mathbf{H}\right)  \right)
=\sum_{k=1}^{2}\mathbb{E}\left[  C\left(  \left\vert H_{j,k}\right\vert
^{2}P_{k}\left(  \mathbf{H}\right)  \right)  \right]  , & j,k=1,2\text{,
}j\not =k
\end{array}
\\%
\begin{array}
[c]{cc}%
S^{\left(  i\right)  }\left(  \underline{P}\left(  \mathbf{H}\right)  \right)
=\mathbb{E}\left[  C\left(  \sum_{k=1}^{2}\left\vert H_{j,k}\right\vert
^{2}P_{k}\left(  \mathbf{H}\right)  \right)  \right]  , & \text{ for }\left(
i,j\right)  =\left(  3a,2\right)  ,\left(  3b,1\right)
\end{array}
\\%
\begin{array}
[c]{cc}%
S^{\left(  3c\right)  }\left(  \underline{P}\left(  \mathbf{H}\right)
\right)  =S^{\left(  3a\right)  }\left(  \underline{P}\left(  \mathbf{H}%
\right)  \right)  , & \text{ }s.t.\text{ }S^{\left(  3a\right)  }\left(
\underline{P}\left(  \mathbf{H}\right)  \right)  =S^{\left(  3b\right)
}\left(  \underline{P}\left(  \mathbf{H}\right)  \right)
\end{array}
\text{ }\\%
\begin{array}
[c]{ccc}%
S^{\left(  l,n\right)  }\left(  \underline{P}\left(  \mathbf{H}\right)
\right)  =S^{\left(  l\right)  }\left(  \underline{P}\left(  \mathbf{H}%
\right)  \right)  , & \text{ }s.t.\text{ }S^{\left(  l\right)  }\left(
\underline{P}\left(  \mathbf{H}\right)  \right)  =S^{\left(  n\right)
}\left(  \underline{P}\left(  \mathbf{H}\right)  \right)  . & \text{for all
}\left(  l,n\right)  .
\end{array}
\end{array}
\label{DF_Jdef}%
\end{equation}
The conditions for each case are (see Figs \ref{Fig_Case12}-\ref{Fig_BC23})
given below where for each case the condition holds true when evaluated at the
optimal policies $\underline{P}^{\left(  i\right)  }(\mathbf{H})$ and
$\underline{P}^{\left(  l,n\right)  }(\mathbf{H})$ for cases $i$ and $\left(
l,n\right)  $, respectively. For ease of notation, we do not explicitly denote
the dependence of $S^{\left(  i\right)  }$ and $S^{\left(  l,n\right)  }$ on
the appropriate $\underline{P}^{\left(  i\right)  }(\mathbf{H})$ and
$\underline{P}^{\left(  l,n\right)  }(\mathbf{H})$, respectively.
\begin{align}
&
\begin{array}
[c]{cl}%
\underline{\text{Case}\ 1}: & S^{\left(  1\right)  }<\min\left(  S^{\left(
3a\right)  },S^{\left(  3b\right)  }\right)
\end{array}
\label{C1Cd}\\
&
\begin{array}
[c]{cl}%
\underline{\text{Case}\ 2}: & S^{\left(  2\right)  }<\min\left(  S^{\left(
3a\right)  },S^{\left(  3b\right)  }\right)
\end{array}
\label{C2Cd}\\
&
\begin{array}
[c]{cc}%
\underline{\text{Case}\ 3a}: & S^{\left(  3a\right)  }<\min\left(  S^{\left(
3b\right)  },S^{\left(  1\right)  },S^{\left(  2\right)  }\right)
\end{array}
\label{C3aCd}\\
&
\begin{array}
[c]{cc}%
\underline{\text{Case}\ 3b}: & S^{\left(  3b\right)  }<\min\left(  S^{\left(
3a\right)  },S^{\left(  1\right)  },S^{\left(  2\right)  }\right)
\end{array}
\label{C3bCd}\\
&
\begin{array}
[c]{cc}%
\underline{\text{Case}\ 3c}: & S^{\left(  3a\right)  }=S^{\left(  3b\right)
}<\min\left(  S^{\left(  1\right)  },S^{\left(  2\right)  }\right)
\end{array}
\label{C3cCd}%
\end{align}

\begin{align}
&
\begin{array}
[c]{cccc}%
\underline{\text{Case}\ \left(  1,3a\right)  }: & S^{\left(  3a\right)
}<S^{\left(  3b\right)  } & \text{and} & S^{\left(  1\right)  }<S^{\left(
3b\right)  }%
\end{array}
\label{C13aCd}\\
&
\begin{array}
[c]{cccc}%
\underline{\text{Case}\ \left(  2,3a\right)  }: & S^{\left(  3a\right)
}<S^{\left(  3b\right)  } & \text{and} & S^{\left(  2\right)  }<S^{\left(
3b\right)  }%
\end{array}
\label{C23aCd}\\
&
\begin{array}
[c]{cccc}%
\underline{\text{Case}\ \left(  1,3b\right)  }: & S^{\left(  3b\right)
}<S^{\left(  3a\right)  } & \text{and} & S^{\left(  1\right)  }<S^{\left(
3a\right)  }%
\end{array}
\label{C13bCd}\\
&
\begin{array}
[c]{cccc}%
\underline{\text{Case}\ \left(  2,3b\right)  }: & S^{\left(  3b\right)
}<S^{\left(  3a\right)  } & \text{and} & S^{\left(  2\right)  }<S^{\left(
3a\right)  }%
\end{array}
\label{C23bCd}\\
&
\begin{array}
[c]{cc}%
\underline{\text{Case}\ \left(  1,3c\right)  }: & S^{\left(  3a\right)
}=S^{\left(  3b\right)  }=S^{\left(  1\right)  }<S^{\left(  2\right)  }%
\end{array}
\label{C13cCd}\\
&
\begin{array}
[c]{cc}%
\underline{\text{Case}\ \left(  2,3c\right)  }: & S^{\left(  3a\right)
}=S^{\left(  3b\right)  }=S^{\left(  2\right)  }<S^{\left(  1\right)  }.
\end{array}
\label{C23cCd}%
\end{align}
\newline

The optimal policy for each case is determined using Lagrange multipliers and
the \textit{Karush}-\textit{Kuhn}-\textit{Tucker} (KKT) conditions. The
sum-capacity optimal \underline{$P$}$^{\ast}\left(  \mathbf{H}\right)  $ is
given by that $\underline{P}^{(i)}\left(  \mathbf{H}\right)  $ or
$\underline{P}^{(l,n)}\left(  \mathbf{H}\right)  $ that satisfies the
conditions of its case in (\ref{C1Cd})-(\ref{C23cCd}).

\begin{remark}
For cases $1$ and $2$, one can expand the capacity expressions to verify that
the conditions $S^{\left(  l\right)  }<\min\left(  S^{\left(  3a\right)
},S^{\left(  3b\right)  }\right)  $, $l=1,2,$ imply that $S^{\left(  1\right)
}<S^{\left(  2\right)  }$ and vice-versa. Therefore, if the optimal policy is
determined in the order of the cases in (\ref{C1Cd})-(\ref{C23cCd}), the
conditions for cases $\left(  1,3c\right)  $ and $\left(  2,3c\right)  $ are
tested only after all other cases have been excluded. Furthermore, the two
cases are mutually exclusive, and thus, (\ref{C13cCd}) and (\ref{C23cCd})
simply redundant conditions written for completeness.
\end{remark}

\begin{remark}
For the two-user case the conditions can be written directly from the geometry
of intersecting rate regions for each case. However, for a more general
$K$-user C-MAC, the conditions can be written using the fact that the rate
regions for any $\underline{P}\left(  \mathbf{H}\right)  $ are polymatroids
and that the sum-rate of two intersecting polymatroids is given by the
polymatroid intersection lemma. A detailed analysis of the rate-region and the
optimal policies using the polymatroid intersection lemma for a $K$-user
two-receiver network is developed in \cite{cap_theorems:LSYLNMHVP}.
\end{remark}

The following theorem summarizes the form of \underline{$P$}$^{\ast}\left(
\mathbf{H}\right)  $ and presents an algorithm to compute it. The optimal
policy maximizing each case can be obtained in a straightforward manner using
standard constrained convex maximization techniques. The algorithm exploits
the fact that each the occurence of one case excludes all other cases and the
case that occurs is the one for which the optimal policy satisfies the case
conditions. We refer the reader to \cite[Appendix]{cap_theorems:LSYLNMHVP} for
a detailed analysis.

\begin{theorem}
\label{Th_CMAC_P}The optimal policy $\underline{P}^{\ast}\left(
\mathbf{H}\right)  $ achieving the sum-capacity of a two-user ergodic fading
C-MAC is obtained by computing $\underline{P}^{(i)}\left(  \mathbf{H}\right)
$ and $\underline{P}^{(l,n)}\left(  \mathbf{H}\right)  $ starting with
cases~$1$ and $2$, followed by cases $3a,$ $3b,$ and $3c$, in that order, and
finally the boundary cases $(l,n),$ in the order that cases $\left(
l,3c\right)  $ are the last to be optimized, until for some case the
corresponding $\underline{P}^{(i)}\left(  \mathbf{H}\right)  $ or
$\underline{P}^{(l,n)}\left(  \mathbf{H}\right)  $ satisfies the case
conditions. The optimal \underline{$P$}$^{\ast}\left(  \mathbf{H}\right)  $ is
given by the optimal $\underline{P}^{(i)}\left(  \mathbf{H}\right)  $ or
$\underline{P}^{(l,n)}\left(  \mathbf{H}\right)  $ that satisfies its case
conditions and falls into one of the following three categories:

\textit{Cases }$1$ \textit{and} $2$: The optimal policies for the two users
are such that each user water-fills over its bottle-neck link, i.e., over the
direct link to that receiver with the smaller (interference-free) ergodic
fading capacity. Thus for cases $1$ and $2$, each transmitter water-fills on
the (interference-free) point-to-point links to its intended and unintended
receivers, respectively. Thus, for case $1$, $P_{k}^{\left(  \ast\right)
}\left(  \mathbf{H}\right)  =P_{k}^{\left(  1\right)  }\left(  \mathbf{H}%
\right)  =P_{k}^{wf}\left(  H_{k,k}\right)  $, and for case 2$,$
$P_{k}^{\left(  \ast\right)  }\left(  \mathbf{H}\right)  =P_{k}^{\left(
2\right)  }\left(  \mathbf{H}\right)  =P_{k}^{wf}\left(  H_{\left\{
1,2\right\}  \backslash k,k}\right)  $, $k=1,2$. where $P_{k}^{wf}\left(
H_{j,k}\right)  $ for $j,k=1,2,$ is defined in Theorem \ref{Th_VS}.

\textit{Cases }$\left(  3a,3b,3c\right)  $: For cases $3a$ and $3b$, the
optimal user policies $P_{k}^{\ast}\left(  \mathbf{H}\right)  $, for all $k$,
are opportunistic multiuser waterfilling solutions over the multiaccess links
to receivers 1 and 2, respectively. For case $3c$, $P_{k}^{\ast}\left(
\mathbf{H}\right)  $, for all $k$, takes an opportunistic non-waterfilling
form and depends on the channel gains for each user at both receivers.

\textit{Boundary Cases}: The optimal user policies $P_{k}^{\ast}\left(
\mathbf{H}\right)  $, for all $k$, are opportunistic non-waterfilling solutions.
\end{theorem}

\begin{remark}
The sum-rate optimal policies for a two-transmitter two-receiver ergodic
fading channel where one of the receiver also acts as a relay is developed in
\cite{cap_theorems:LSYLNMHVP}. The analysis here is very similar to that in
\cite{cap_theorems:LSYLNMHVP}, and thus, we briefly outline the intuition
behind the results in the proof below.
\end{remark}

\begin{proof}
The optimal policy for each case can be determined in a straightforward manner
using Lagrange multipliers and the \textit{Karush}-\textit{Kuhn}%
-\textit{Tucker} (KKT) conditions. Furthermore, not including all or some of
the constraints for each case in the maximization problem simplifies the
determination of the solution.

For cases $1$ and $2$, $S^{\left(  1\right)  }$ and $S^{\left(  2\right)  }$,
respectively, are sum of two bottle-neck point-to-point links, and thus, are
maximized by the single-user waterfilling power policies, one for each
bottle-neck link. For cases $3a$ and $3b$, the optimization is equivalent to
maximizing the sum-capacity at one of the receivers. Thus, applying the
results in \cite[Lemma 3.10]{cap_theorems:TH01} (see also
\cite{cap_theorems:Knopp_Humblet}), for these two cases, one can show that
sum-capacity achieving policies are opportunistic waterfilling solutions that
exploit the multiuser diversity.

For case $3c$, the sum-rate $S^{\left(  3a\right)  }$ is maximized subject to
the constraint $S^{\left(  3a\right)  }=S^{\left(  3b\right)  }$. Thus, for
this case, the KKT conditions can be used to show that while opportunistic
scheduling of the users based on a function of their fading states to both
receivers is optimal, the optimal policies are no longer waterfilling
solutions. The same argument also holds for the boundary cases $\left(
l,n\right)  $ where $S^{\left(  l\right)  }$ is maximized subject to
$S^{\left(  l\right)  }=S^{\left(  n\right)  }$. In all cases, the optimal
policies can be determined using an iterative procedure in a manner akin to
the iterative waterfilling approach for fading MACs
\cite{cap_theorems:Yu_Rhee02}. See \cite[Appendix]{cap_theorems:LSYLNMHVP}%
\ for a detailed proof.
\end{proof}

\subsection{Capacity Region:\ Optimal Policies}

As mentioned earlier, each point on the boundary of $\mathcal{C}%
_{\text{C-MAC}}\left(  \overline{P}_{1},\overline{P}_{2}\right)  $ is obtained
by maximizing the weighted sum $\mu_{1}R_{1}$ $+$ $\mu_{2}R_{2}$ over all
$\underline{P}\left(  \mathbf{H}\right)  \in\mathcal{P}$, and for all $\mu
_{1}>0$, $\mu_{2}>0$, subject to (\ref{CMAC_OB}). Without loss of generality,
we assume that $\mu_{1}<\mu_{2}$. Let \underline{$\mu$} denote the pair
$\left(  \mu_{1},\mu_{2}\right)  $. The optimal policy $\underline{P}^{\ast
}\left(  \mathbf{H,}\underline{\mu}\right)  $ is given by%
\begin{equation}
\underline{P}^{\ast}\left(  \mathbf{H,}\underline{\mu}\right)  =\arg
\max_{\underline{P}\in\mathcal{P}}\left(  \mu_{1}R_{1}+\mu_{2}R_{2}\right)
\text{ s.t. }\left(  R_{1},R_{2}\right)  \in\mathcal{C}_{\text{C-MAC}}\left(
\overline{P}_{1},\overline{P}_{2}\right)  \label{CMAC_CapR}%
\end{equation}
where $\mu_{1}R_{1}+\mu_{2}R_{2}$, denoted by $S^{\left(  x\right)  }\left(
\underline{\mu},\underline{P}\left(  \mathbf{H}\right)  \right)  $ for case
$x=i,\left(  l,n\right)  $, for all $i$ and $\left(  l,n\right)  $, for the
different cases are given by%
\begin{equation}%
\begin{array}
[c]{l}%
S^{\left(  1\right)  }\left(  \underline{\mu},\underline{P}\left(
\mathbf{H}\right)  \right)  =\sum_{k=1}^{2}\mu_{k}\mathbb{E}\left[  C\left(
\left\vert H_{k,k}\right\vert ^{2}P_{k}\left(  \mathbf{H}\right)  \right)
\right] \\%
\begin{array}
[c]{cc}%
S^{\left(  2\right)  }\left(  \underline{\mu},\underline{P}\left(
\mathbf{H}\right)  \right)  =\sum_{k=1}^{2}\mu_{k}\mathbb{E}\left[  C\left(
\left\vert H_{j,k}\right\vert ^{2}P_{k}\left(  \mathbf{H}\right)  \right)
\right]  , & j,k=1,2\text{, }j\not =k
\end{array}
\\%
\begin{array}
[c]{cc}%
S^{\left(  i\right)  }\left(  \underline{\mu},\underline{P}\left(
\mathbf{H}\right)  \right)  =\mu_{1}S^{\left(  i\right)  }\left(
\underline{P}\left(  \mathbf{H}\right)  \right)  +\left(  \mu_{2}-\mu
_{1}\right)  \min\limits_{j=1,2}\left(  \mathbb{E}\left[  C\left(  \left\vert
H_{j,2}\right\vert ^{2}P_{2}\left(  \mathbf{H}\right)  \right)  \right]
\right)  & i=3a,3b
\end{array}
\\%
\begin{array}
[c]{cc}%
S^{\left(  3c\right)  }\left(  \underline{\mu},\underline{P}\left(
\mathbf{H}\right)  \right)  =S^{\left(  3a\right)  }\left(  \underline
{P}\left(  \mathbf{H}\right)  \right)  , & \text{ }s.t.\text{ }S^{\left(
3a\right)  }\left(  \underline{\mu},\underline{P}\left(  \mathbf{H}\right)
\right)  =S^{\left(  3b\right)  }\left(  \underline{\mu},\underline{P}\left(
\mathbf{H}\right)  \right)
\end{array}
\text{ }\\%
\begin{array}
[c]{ccc}%
S^{\left(  l,n\right)  }\left(  \underline{\mu},\underline{P}\left(
\mathbf{H}\right)  \right)  =S^{\left(  l\right)  }\left(  \underline
{P}\left(  \mathbf{H}\right)  \right)  , & \text{ }s.t.\text{ }S^{\left(
l\right)  }\left(  \underline{\mu},\underline{P}\left(  \mathbf{H}\right)
\right)  =S^{\left(  n\right)  }\left(  \underline{\mu},\underline{P}\left(
\mathbf{H}\right)  \right)  . & \text{for all }\left(  l,n\right)  .
\end{array}
\end{array}
\label{CMAC_SR}%
\end{equation}
The expressions for $\mu_{2}<\mu_{1}$ can be obtained from (\ref{CMAC_SR}) by
interchanging the indexes $1$ and $2$ in the second term in the expression for
$S^{\left(  i\right)  }\left(  \underline{\mu},\underline{P}\left(
\mathbf{H}\right)  \right)  $, $i=3a,$ $3b$. From the convexity of
$\mathcal{C}_{\text{C-MAC}}$, every point on the boundary is obtained from the
intersection of two MAC rate regions. From Figs. \ref{Fig_Case12}%
-\ref{Fig_BC23}, we see that for cases $1$, $2$, and the boundary cases, the
region of intersection has a unique vertex at which both user rates are
non-zero and thus, $\mu_{1}R_{1}+\mu_{2}R_{2}$ will be tangential to that
vertex. On the other hand, for cases $3a$, $3b$, and $3c$, the intersecting
region is also a pentagon and thus, $\mu_{1}R_{1}+\mu_{2}R_{2}$, for $\mu
_{1}<\mu_{2}$, is maximized by that vertex at which user $2$ is decoded after
user $1$. The conditions for the different cases are given by (\ref{C1Cd}%
)-(\ref{C23cCd}). Note that for case $1$, since the sum-capacity achieving
policies also achieve the point-to-point link capacities for each user to its
intended destination, the capacity region is simply given by the single-user
capacity bounds on $R_{1}$ and $R_{2}$.

The following theorem summarizes the capacity region of an ergodic fading
C-MAC and the optimal policies that achieve it for $\mu_{1}<\mu_{2}$. The
policies for $\mu_{1}>\mu_{2}$ can be obtained in a straightforward manner.

\begin{theorem}
\label{Th_CMAC_CR}The optimal policy $\underline{P}^{\ast}\left(
\mathbf{H}\right)  $ achieving the sum-capacity of a two-user ergodic fading
C-MAC is obtained by computing $\underline{P}^{(i)}\left(  \mathbf{H}\right)
$ and $\underline{P}^{(l,n)}\left(  \mathbf{H}\right)  $ starting with the
inactive cases~$1$ and $2$, followed by the active cases $3a,$ $3b,$ and $3c$,
in that order, and finally the boundary cases $(l,n),$ in the order that cases
$\left(  l,3c\right)  $ are the last to be optimized, until for some case the
corresponding $\underline{P}^{(i)}\left(  \mathbf{H}\right)  $ or
$\underline{P}^{(l,n)}\left(  \mathbf{H}\right)  $ satisfies the case
conditions. The optimal \underline{$P$}$^{\ast}\left(  \mathbf{H}\right)  $ is
given by the optimal $\underline{P}^{(i)}\left(  \mathbf{H}\right)  $ or
$\underline{P}^{(l,n)}\left(  \mathbf{H}\right)  $ that satisfies its case
conditions and falls into one of the following three categories:

\textit{Inactive Cases}: The optimal policies for the two users are such that
each user water-fills over its bottle-neck link. Thus for cases $1$ and $2$,
each transmitter water-fills on the (interference-free) point-to-point links
to its intended and unintended receivers, respectively. Thus, for case $1$,
$P_{k}^{\left(  \ast\right)  }\left(  \mathbf{H}\right)  =P_{k}^{wf}\left(
H_{k,k}\right)  $, and for case 2$,$ $P_{k}^{\left(  \ast\right)  }\left(
\mathbf{H}\right)  =P_{k}^{\left(  2\right)  }\left(  \mathbf{H}\right)
=\mu_{k}P_{k}^{wf}\left(  H_{\left\{  1,2\right\}  \backslash k,k}\right)  $,
$k=1,2$, where $P_{k}^{wf}\left(  H_{j,k}\right)  $ for $j,k=1,2,$ is defined
in Theorem \ref{Th_VS}.

\textit{Cases }$\left(  3a,3b,3c\right)  $: For cases $3a$ and $3b$, the
optimal policies are opportunistic multiuser solutions given in for the
special case where the minimum sum-rate and single-user rate for user $2$ are
achieved at the same receiver. Otherwise, the solutions for all three cases
are opportunistic non-waterfilling solutions.

\textit{Boundary Cases}: The optimal policies maximizing the constrained
optimization of $S_{\mu_{1},\mu_{2}}^{\left(  l,n\right)  }\left(
\underline{P}\left(  \mathbf{H}\right)  \right)  $ are also opportunistic
non-waterfilling solutions.
\end{theorem}

\section{\label{Sec_4}Proofs}

\subsection{Ergodic VS IFCs: Proof of Theorem \ref{Th_VS}}

We now prove Theorem \ref{Th_VS} on the sum-capacity of a sub-class of ergodic
fading IFCs with a mix of weak and strong sub-channels. The capacity achieving
scheme requires both receivers to decode both messages.

\subsubsection{Converse}

An outer bound on the sum-capacity of an interference channel is given by the
sum-capacity of a IFC in which interference has been eliminated at one or both
receivers. One can view it alternately as providing each receiver with the
codeword of the interfering transmitter. Thus, from\ Fano's and the data
processing inequalities we have that the achievable rate must satisfy
\begin{subequations}
\begin{align}
n\left(  R_{1}+R_{2}\right)  -n\epsilon &  \leq I(X_{1}^{n};Y_{1}^{n}%
|X_{2}^{n},\mathbf{H}^{n})+I(X_{2}^{n};Y_{2}^{n}|X_{1}^{n},\mathbf{H}^{n})\\
&  =I(X_{1}^{n};\tilde{Y}_{1}^{n}|\mathbf{H}^{n})+I(X_{2}^{n};\tilde{Y}%
_{2}^{n}|\mathbf{H}^{n}) \label{VS_OB2}%
\end{align}
where
\end{subequations}
\begin{equation}%
\begin{array}
[c]{cc}%
\tilde{Y}_{k}=H_{k,k}X_{k}+Z_{k}, & k=1,2.
\end{array}
\label{YX_P2P}%
\end{equation}

The converse proof techniques developed in \cite[Appendix]%
{cap_theorems:GoldsmithVaraiya} for a point-to-point ergodic fading link in
which the transmit and received signals are related by (\ref{YX_P2P}) can be
apply directly following (\ref{VS_OB2}), and thus, we have that any achievable
rate pair must satisfy
\begin{equation}
R_{1}+R_{2}\leq\sum_{k=1}^{2}\mathbb{E}\left[  C\left(  \left\vert
H_{k,k}\right\vert ^{2}P_{k}^{wf}\left(  H_{k,k}\right)  \right)  \right]  .
\label{VS_OB}%
\end{equation}

\subsubsection{Achievable Scheme}

Corollary \ref{Cor_1} states that the capacity region of an equivalent C-MAC
is an inner bound on the capacity region of an IFC. Thus, from Theorem
\ref{Th_CMAC_P} a sum-rate of
\begin{equation}
\sum_{k=1}^{2}\mathbb{E}\left[  C\left(  \left\vert H_{k,k}\right\vert
^{2}P_{k}^{wf}\left(  H_{k,k}\right)  \right)  \right]  \label{EVS_SR}%
\end{equation}
is achievable when $\underline{P}^{\ast}\left(  \mathbf{H}\right)
=\underline{P}^{wf}\left(  H_{k,k}\right)  $ satisfies the condition for case
1 in (\ref{C1Cd}), which is equivalent to the requirement that $\underline
{P}^{wf}\left(  H_{k,k}\right)  $ satisfies (\ref{EVS_Cond}).

The conditions in (\ref{EVS_Cond}) imply that waterfilling over the two
point-to-point links from each user to its receiver is optimal when the fading
averaged rate achieved by each transmitter at its intended receiver is
strictly smaller than the rate it achieves in the presence of interference at
the unintended receiver, i.e., the channel is very strong on average.

Finally, since the achievable bound on the sum-rate in (\ref{EVS_SR}) also
achieves the single-user capacities, the capacity region of an EVS IFC is
given by (\ref{EVS_CapR}).

\subsubsection{Separability}

Achieving the sum-capacity and the capacity region of the C-MAC requires joint
encoding and decoding across all sub-channels. This observation also carries
over to the sub-class of ergodic very strong IFCs that are in general a mix of
weak and strong sub-channels. In fact, any strategy where each sub-channel is
viewed as an independent IFC will be strictly sub-optimal except for those
cases where every sub-channel is very strong at the optimal policy.

\subsection{Uniformly Strong IFC: Proof of Theorem \ref{Th_Str}}

We now show that the strategy of allowing both receivers to decode both
messages achieves the sum-capacity for the sub-class of fading IFCs in which
every fading state (sub-channel) is strong, i.e., the entries of $\mathbf{h}$
satisfy $\left\vert h_{1,1}\right\vert <\left\vert h_{2,1}\right\vert $ and
$\left\vert h_{2,2}\right\vert <\left\vert h_{1,2}\right\vert $.

\subsubsection{Converse}

In the Proof of Theorem \ref{Th_VS}, we developed a genie-aided outer bound on
the sum-capacity of ergodic fading IFCs. One can use similar arguments to
write the bounds on the rates $R_{1}$ and $R_{2}$, for every choice of
feasible power policy $\underline{P}\left(  \mathbf{H}\right)  $, as
\begin{align}
R_{k}  &  \leq\mathbb{E}\left[  \log\left(  1+\left\vert H_{k,k}\right\vert
^{2}P_{k}\left(  \mathbf{H}\right)  \right)  \right]  ,\text{ }%
k=1,2.\label{US_SU_Rate}\\
\text{ \ \ \ }  &  \leq\mathbb{E}\left[  \log\left(  1+\left\vert
H_{j,k}\right\vert ^{2}P_{k}\left(  \mathbf{H}\right)  \right)  \right]
,\text{ \ }j=1,2,j\not =k, \label{US_SU_Rate2}%
\end{align}
where (\ref{US_SU_Rate2}) follows from the uniformly strong condition in
(\ref{US_HCond}). We now present two additional bounds where the genie reveals
the interfering signal to only one of the receivers. Consider first the case
where the genie reveals the interfering signal at receiver $2.$ One can then
reduce the two-sided IFC to a one-sided IFC, i.e., set $H_{2,1}=0$.

For this genie-aided one-sided channel, from Fano's inequality, we have that
the achievable rate must satisfy
\begin{subequations}
\begin{equation}
n\left(  R_{1}+R_{2}\right)  -n\epsilon\leq I(X_{1}^{n};Y_{1}^{n}%
|\mathbf{H}^{n})+I(X_{2}^{n};Y_{2}^{n}|\mathbf{H}^{n}). \label{USFano}%
\end{equation}
We first consider the expression on the right-side of (\ref{USFano}) for some
instantiation $\mathbf{h}^{n}$. We thus have
\end{subequations}
\begin{equation}
I(X_{1}^{n};Y_{1}^{n}|\mathbf{H}^{n}=\mathbf{h}^{n})+I(X_{2}^{n};Y_{2}%
^{n}|\mathbf{H}^{n}=\mathbf{h}^{n})=I(X_{1}^{n};\mathbf{h}_{1,1}^{n}X_{1}%
^{n}+\mathbf{h}_{1,2}^{n}X_{2}^{n}+Z_{1}^{n})+I(X_{2}^{n};\mathbf{h}_{2,2}%
^{n}X_{2}^{n}+Z_{2}^{n}) \label{StrCon3}%
\end{equation}
where $\mathbf{h}_{j,k}^{n}$ is a diagonal matrix with diagonal entries
$h_{j,k,i}$, for all $i=1,2,\ldots,n$. Consider the mutual information terms
on the right-side of the equality in (\ref{StrCon3}). We can expand these
terms as
\begin{subequations}
\begin{align}
&  h\left(  \mathbf{h}_{1,1}^{n}X_{1}^{n}+\mathbf{h}_{1,2}^{n}X_{2}^{n}%
+Z_{1}^{n}\right)  -h\left(  \mathbf{h}_{1,2}^{n}X_{2}^{n}+Z_{1}^{n}\right) \\
&  +h\left(  \mathbf{h}_{2,2}^{n}X_{2}^{n}+Z_{2}^{n}\right)  -h\left(
Z_{2}^{n}\right) \nonumber\\
&  \overset{\left(  a\right)  }{\leq}n\sum_{i=1}^{n}(h\left(  h_{1,1,i}%
X_{1,i}+h_{1,2,i}X_{2,i}+Z_{1,i}\right)  -h\left(  Z_{2,i}\right)
)\label{US_Con2}\\
&  -h\left(  h_{1,2}^{n}X_{2}^{n}+Z_{1}^{n}\right)  +h\left(  h_{2,2}^{n}%
X_{2}^{n}+Z_{2}^{n}\right)  ,
\end{align}
where $\left(  a\right)  $ is from the fact that conditioning does not
increase entropy. For the uniformly strong ergodic IFC satisfying
(\ref{US_HCond}), i.e., $\left\vert h_{2,2,i}\right\vert \leq\left\vert
h_{1,2,i}\right\vert ,$ for all $i=1,2,\ldots,n,$ the third and fourth terms
in (\ref{US_Con2}) can be simplified as
\end{subequations}
\begin{subequations}
\label{USConExp}%
\begin{align}
&  -h\left(  X_{2}^{n}+\left(  \mathbf{h}_{1,2}^{n}\right)  ^{-1}Z_{1}%
^{n}\right)  +h\left(  X_{2}^{n}+\left(  \mathbf{h}_{2,2}^{n}\right)
^{-1}Z_{2}^{n}\right) \\
&  -\log\left(  \left\vert \mathbf{h}_{1,2}^{n}\right\vert \right)
+\log\left(  \left\vert \mathbf{h}_{2,2}^{n}\right\vert \right) \nonumber\\
&  =-h\left(  X_{2}^{n}+\left(  \mathbf{h}_{1,2}^{n}\right)  ^{-1}Z_{1}%
^{n}\right)  +h\left(  X_{2}^{n}+\left(  \mathbf{h}_{1,2}^{n}\right)
^{-1}Z_{1}^{n}+\tilde{Z}^{n}\right) \\
&  -\log\left(  \left\vert \mathbf{h}_{1,2}^{n}\right\vert \right)
+\log\left(  \left\vert \mathbf{h}_{2,2}^{n}\right\vert \right) \nonumber\\
&  =I(\tilde{Z}^{n};X_{2}^{n}+\left(  \mathbf{h}_{1,2}^{n}\right)  ^{-1}%
Z_{1}^{n}+\tilde{Z}^{n})-\log\left(  \left\vert \mathbf{h}_{1,2}%
^{n}\right\vert \right)  +\log\left(  \left\vert \mathbf{h}_{2,2}%
^{n}\right\vert \right) \\
&  \leq I(\tilde{Z}^{n};\left(  \mathbf{h}_{1,2}^{n}\right)  ^{-1}Z_{1}%
^{n}+\tilde{Z}^{n})-\log\left(  \left\vert \mathbf{h}_{1,2}^{n}\right\vert
\right)  +\log\left(  \left\vert \mathbf{h}_{2,2}^{n}\right\vert \right) \\
&  =h(Z_{2}^{n})-h(Z_{1}^{n})\label{US_Hsimp}\\
&  =\sum_{i=1}^{n}\left(  h(Z_{2,i})-h(Z_{1,i})\right)
\end{align}
where $\tilde{Z}_{i}\sim\mathcal{CN}\left(  0,\left\vert h_{2,2,i}%
^{-1}\right\vert ^{2}-\left\vert h_{1,2,i}^{-1}\right\vert ^{2}\right)  $, for
all $i$, and the inequality in (\ref{USConExp}) results from the fact that
mixing increases entropy.

Substituting (\ref{US_Hsimp}) in (\ref{US_Con2}), we thus have that for every
instantiation, the $n$-letter expressions reduce to a sum of single-letter
expressions. Over all fading instantiations, one can thus write%
\end{subequations}
\begin{equation}
\left(  R_{1}+R_{2}\right)  -\epsilon\leq I(X_{1}\left(  Q\left(  n\right)
\right)  X_{2}\left(  Q\left(  n\right)  \right)  ;Y_{1}\left(  Q\left(
n\right)  \right)  |\mathbf{H}\left(  Q\left(  n\right)  \right)  Q\left(
n\right)  )
\end{equation}
where $Q\left(  n\right)  $ is a random variable distributed uniformly on
$\left\{  1,2,\ldots,n\right\}  $.

Our analysis from here on is exactly similar to that for a fading MAC in
\cite[Appendix A]{cap_theorems:TH01}, and thus, we omit it in the interest of
space. Effectively, the analysis involves considering an increasing sequence
of partitions (quantized ranges) $I_{k},$ $k=\mathcal{I}^{+}$, of the alphabet
of $\mathbf{H}$, while ensuring that for each $k$, the transmitted signals are
constrained in power. Taking limits appropriately over $n$ and $k$, as in
\cite[Appendix A]{cap_theorems:TH01}, we obtain%

\begin{equation}
R_{1}+R_{2}-\epsilon\leq\mathbb{E}\left[  C\left(
{\textstyle\sum\nolimits_{k=1}^{2}}
\left\vert H_{1,k}\right\vert ^{2}P_{k}\left(  \mathbf{H}\right)  \right)
\right]  \label{US_ConFin4}%
\end{equation}
\qquad where $P(\mathbf{H})$ satisfies (\ref{ErgPwr}).

One can similarly let $H_{1,2}=0$ and show that%
\begin{equation}
R_{1}+R_{2}-\epsilon\leq\mathbb{E}\left[  C\left(
{\textstyle\sum\nolimits_{k=1}^{2}}
\left\vert H_{2,k}\right\vert ^{2}P_{k}\left(  \mathbf{H}\right)  \right)
\right]  \label{US_ConFin5}%
\end{equation}
Combining (\ref{US_SU_Rate}), (\ref{US_SU_Rate2}), (\ref{US_ConFin4}), and
(\ref{US_ConFin5}), we see that, for every choice of $\underline{P}\left(
\mathbf{H}\right)  $, the capacity region of a uniformly strong ergodic fading
IFC lies within the capacity region of a C-MAC for which the fading states
satisfy (\ref{US_HCond}). Thus, over all power policies, we have
\begin{equation}
\mathcal{C}_{\text{IFC}}\left(  \overline{P}_{1},\overline{P}_{2}\right)
\subseteq\mathcal{C}_{\text{C-MAC}}\left(  \overline{P}_{1},\overline{P}%
_{2}\right)  .
\end{equation}

\subsubsection{Achievable Strategy}

Allowing both receivers to decode both messages as stated in Corollary
\ref{Cor_1} achieves the outer bound. For the resulting C-MAC, the uniformly
strong condition in (\ref{US_HCond}) limits the intersection of the rate
regions $\mathcal{C}_{1}\left(  \underline{P}\left(  \mathbf{H}\right)
\right)  $ and $\mathcal{C}_{2}\left(  \underline{P}\left(  \mathbf{H}\right)
\right)  $, for any choice of $\underline{P}\left(  \mathbf{H}\right)  $, to
one of cases $1,$ $3a$, $3b$, $3c$, or the boundary cases $\left(  1,n\right)
$ for $n=3a,3b,3c,$ such that (\ref{US_SU_Rate}) defines the single-user rate bounds.

The sum-capacity optimal policy for each of the above cases is given by
Theorem \ref{Th_CMAC_P}. Thus, the optimal user policies are single-user
waterfilling solutions when the uniformly strong fading IFC also satisfies
(\ref{EVS_Cond}), i.e., the optimal policies satisfy the conditions for case
$1$. For all other cases, the optimal policies are opportunistic multiuser
allocations. Specifically, cases $3a$ and $3b$ the solutions are the classical
multiuser waterfilling solutions \cite{cap_theorems:TH01}.

One can similarly develop the optimal policies that achieve the capacity
region. Here too, for every point $\mu_{1}R_{1}+\mu_{2}R_{2}$, $\mu_{1}%
,\mu_{2},$ on the boundary of the capacity region, the optimal policy
$\underline{P}^{\ast}\left(  \mathbf{H}\right)  $ is either $\underline
{P}^{\left(  1\right)  }\left(  \mathbf{H}\right)  $ or $\underline
{P}^{\left(  n\right)  }\left(  \mathbf{H}\right)  $ or $\underline
{P}^{\left(  1,n\right)  }\left(  \mathbf{H}\right)  $ for $n=3a,$ $3b,$ $3c$.

\subsubsection{Separability}

See Remark \ref{Rem_USSep}.

\subsection{Uniformly Weak One-Sided IFC: Proof of Theorem \ref{Th_UW1}}

We now prove Theorem \ref{Th_UW1} on the sum-capacity of a sub-class of
one-sided ergodic fading IFCs where every sub-channel is weak, i.e., the
channel is uniformly weak. We show that it is optimal to ignore the
interference at the unintended receiver.

\subsubsection{Converse}

From\ Fano's inequality, any achievable rate pair $\left(  R_{1},R_{2}\right)
$ must satisfy
\begin{subequations}
\begin{equation}
n\left(  R_{1}+R_{2}\right)  -n\epsilon\leq I(X_{1}^{n};Y_{1}^{n}%
|\mathbf{H}^{n})+I(X_{2}^{n};Y_{2}^{n}|\mathbf{H}^{n}). \label{UW_Fano}%
\end{equation}
We first consider the expression on the right-side of (\ref{UW_Fano}) for some
instantiation $\mathbf{h}^{n}$, i.e., consider
\end{subequations}
\begin{equation}
I(X_{1}^{n};Y_{1}^{n}|\mathbf{H}^{n}=\mathbf{h}^{n})+I(X_{2}^{n};Y_{2}%
^{n}|\mathbf{H}^{n}=\mathbf{h}^{n})=I(X_{1}^{n};\mathbf{h}_{1,1}^{n}X_{1}%
^{n}+\mathbf{h}_{1,2}^{n}X_{2}^{n}+Z_{1}^{n})+I(X_{2}^{n};\mathbf{h}_{2,2}%
^{n}X_{2}^{n}+Z_{2}^{n}) \label{Con_W1}%
\end{equation}
where $\mathbf{h}_{j,k}^{n}$ is a diagonal matrix with diagonal entries
$h_{j,k,i}$, for all $i=1,2,\ldots,n$. Let $N^{n}$ be a sequence of
independent Gaussian random variables, such that
\begin{equation}
\left[
\begin{array}
[c]{c}%
Z_{1,i}\\
N_{i}%
\end{array}
\right]  \sim\mathcal{CN}\left(  0,\left[
\begin{array}
[c]{cc}%
1 & \rho_{i}\sigma_{i}\\
\rho_{i}\sigma_{i} & \sigma_{i}^{2}%
\end{array}
\right]  \right)  , \label{ConW_ZN}%
\end{equation}
and
\begin{align}
\rho_{i}^{2}  &  =1-\left(  \left\vert h_{1,2,i}\right\vert ^{2}\left/
\left\vert h_{2,2,i}\right\vert ^{2}\right.  \right) \label{ConW_rho}\\
\rho_{i}\sigma_{i}  &  =1+\left\vert h_{2,2,i}\right\vert ^{2}P_{2,i}.
\label{ConW_rs}%
\end{align}
We bound (\ref{Con_W1}) as follows:%
\begin{subequations}
\begin{align}
&  I(X_{1}^{n};Y_{1}^{n}|\mathbf{h}^{n})+I(X_{2}^{n};Y_{2}^{n}|\mathbf{h}%
^{n})\nonumber\\
&  \leq I(X_{1}^{n};Y_{1}^{n},h_{1,1}^{n}X_{1}^{n}+N^{n}|\mathbf{h}%
^{n})+I(X_{2}^{n};Y_{2}^{n}|\mathbf{h}^{n})\\
&  =h\left(  h_{2,2}^{n}X_{2}^{n}+Z_{2}^{n}\right)  -h\left(  Z_{2}%
^{n}\right)  +h\left(  h_{1,1}^{n}X_{1}^{n}+N^{n}\right)  -h\left(
N^{n}\right) \\
&  +h\left(  h_{1,1}^{n}X_{1}^{n}+h_{1,2}^{n}X_{2}^{n}+Z_{1}^{n}|h_{1,1}%
^{n}X_{1}^{n}+N^{n}\right)  -h\left(  h_{1,2}^{n}X_{2}^{n}+Z_{1}^{n}%
|N^{n}\right) \nonumber\\
&  \leq\sum\limits_{i=1}^{n}h\left(  h_{1,1,i}X_{1,i}^{\ast}+N_{i}\right)
-\sum\limits_{i=1}^{n}h\left(  Z_{2,i}\right)  -\sum\limits_{i=1}^{n}h\left(
N_{i}\right)  +h\left(  h_{2,2}^{n}X_{2}^{n}+Z_{2}^{n}\right)  \label{Con_WG}%
\\
&  -h\left(  h_{1,2}^{n}X_{2}^{n}+Z_{1}^{n}|N^{n}\right)  +\sum\limits_{i=1}%
^{n}h\left(  h_{1,1,i}X_{1,i}^{\ast}+h_{1,2,i}X_{2,i}^{\ast}+Z_{1,i}%
|h_{1,1,i}X_{1,i}^{\ast}+N_{i}\right) \nonumber\\
&  =\sum\limits_{i=1}^{n}\left\{  h\left(  h_{1,1,i}X_{1,i}^{\ast}%
+N_{i}\right)  -h\left(  Z_{2,i}\right)  -h\left(  N_{i}\right)  \right.
+h\left(  h_{2,2,i}X_{2,i}^{\ast}+Z_{2,i}\right) \label{Con_WG2}\\
&  -h\left(  h_{1,2,i}X_{2,i}^{\ast}+Z_{1,i}|N_{i}\right)  \left.  +h\left(
h_{1,1,i}X_{1,i}^{\ast}+h_{1,2,i}X_{2,i}^{\ast}+Z_{1,i}|h_{1,1,i}X_{1,i}%
^{\ast}+N_{i}\right)  \right\} \nonumber\\
&  =\sum\limits_{i=1}^{n}\left\{  \log\left(  \left\vert h_{1,1,i}\right\vert
^{2}P_{1,i}+\sigma_{i}^{2}\right)  -h\left(  \sigma_{i}\right)  \right.
+\log\left(  \left\vert h_{2,2,i}\right\vert ^{2}P_{2,i}+1\right)
\label{Con_WG3}\\
&  -\log\left(  \left\vert h_{1,2,i}\right\vert ^{2}P_{2,i}+\left(  1-\rho
_{i}^{2}\right)  \right)  +\log\left(  \left\vert h_{1,1,i}\right\vert
^{2}P_{1,i}+\left\vert h_{1,2,i}\right\vert ^{2}P_{2,i}+1\right. \nonumber\\
&  \left.  \left.  -\left(  \left\vert h_{1,1,i}\right\vert ^{2}P_{1,i}%
+\sigma_{i}\right)  ^{-1}\left(  \left\vert h_{1,1,i}\right\vert ^{2}%
P_{1,i}+\rho_{i}\sigma_{i}\right)  ^{2}\right)  \right\} \nonumber\\
&  =\sum\limits_{i=1}^{n}\left\{  \log\left(  \left\vert h_{2,2,i}\right\vert
^{2}P_{2,i}+1\right)  +\log\left(  1+\frac{\left\vert h_{1,1,i}\right\vert
^{2}P_{1,i}}{1+\left\vert h_{1,2,i}\right\vert ^{2}P_{2,i}}\right)  \right\}
\label{Con_WG4}%
\end{align}
where (\ref{Con_WG}) follows from the fact that conditioning does not increase
entropy and that the conditional entropy is maximized by Gaussian signaling,
i.e., $X_{k,i}^{\ast}\sim\mathcal{CN}(0,P_{k,i})$, (\ref{Con_WG2}) follows
from (\ref{ConW_ZN}) and (\ref{ConW_rho}) which imply%
\end{subequations}
\begin{equation}
var\left(  h_{1,2,i}^{-1}Z_{1,i}|N_{i}\right)  =\frac{1-\rho_{i}^{2}%
}{\left\vert h_{1,2,i}\right\vert ^{2}}=\left\vert h_{2,2,i}\right\vert ^{-2}%
\end{equation}
and therefore,
\begin{subequations}
\begin{align}
&  h\left(  h_{2,2}^{n}X_{2}^{n}+Z_{2}^{n}\right)  -h\left(  h_{1,2}^{n}%
X_{2}^{n}+Z_{1}^{n}|N^{n}\right) \\
&  =\log\left(  \left\vert h_{2,2}^{n}\right\vert \right)  -\log\left(
\left\vert h_{1,2}^{n}\right\vert \right) \\
&  =\sum\limits_{i=1}^{n}h\left(  h_{2,2,i}X_{2,i}^{\ast}+Z_{2,i}\right)
-h\left(  h_{1,2,i}X_{2,i}^{\ast}+Z_{1,i}|N_{i}\right)  ; \label{ConW_2terms}%
\end{align}
and (\ref{Con_WG4}) follows from substituting (\ref{ConW_rs}) in
(\ref{Con_WG3}) and simplifying the resulting expressions.

Our analysis from here on is similar to that for the US IFC (see also
\cite[Appendix A]{cap_theorems:TH01}). Effectively, the analysis involves
considering an increasing sequence of partitions (quantized ranges) $I_{k},$
$k=\mathcal{I}^{+}$, of the alphabet of $\mathbf{H}$, while ensuring that for
each $k$, the transmitted signals are constrained in power. Taking limits
appropriately over $n$ and $k$, and using the fact that the $\log$ expressions
in (\ref{Con_WG4}) are concave functions of $P_{k,i}$, for all $k$, and that
every feasible power policy satisfies (\ref{ErgPwr}), we obtain%
\end{subequations}
\begin{subequations}
\begin{equation}
R_{1}+R_{2}-\epsilon\leq\mathbb{E}\left[  C\left(  \left\vert H_{2,2}%
\right\vert ^{2}P_{2}\left(  \mathbf{h}\right)  \right)  +C\left(
\frac{\left\vert H_{1,1}\right\vert ^{2}P_{1}\left(  \mathbf{h}\right)
}{1+\left\vert H_{1,2}\right\vert ^{2}P_{2}\left(  \mathbf{h}\right)
}\right)  \right]  . \label{ConW_fin}%
\end{equation}
An outer bound on the sum-rate is obtained by maximizing over all feasible
policies and is given by (\ref{SC_Weak}) and (\ref{SCW_S}).

\subsubsection{Achievable Strategy}

The outer bounds can be achieved by letting receiver 1 ignore (not decode) the
interference it sees from transmitter $2$. Averaged over all sub-channels, the
sum of the rates achieved at the two receivers for every choice of
$\underline{P}\left(  \mathbf{H}\right)  $ is given by (\ref{ConW_fin}). The
sum-capacity in (\ref{SC_Weak}) is then obtained by maximizing (\ref{ConW_fin}%
) over all feasible $\underline{P}\left(  \mathbf{H}\right)  $.

\subsubsection{Separability}

The optimality of separate encoding and decoding across the sub-channels
follows directly from the fact that the sub-channels are all of the same type,
and thus, independent messages can be multiplexed across the sub-channels.
This is in contrast to the uniformly strong and the ergodic very strong IFCs
where mixtures of different channel types in both cases is exploited to
achieve the sum-capacity by encoding and decoding jointly across all sub-channels.
\end{subequations}
\begin{remark}
A natural question is whether one can extend the techniques developed here to
the two-sided UW IFC. In this case, one would have four parameters per channel
state, namely $\rho_{k}\left(  \mathbf{H}\right)  $ and $\sigma_{k}^{2}\left(
\mathbf{H}\right)  $, $k=1,2$. Thus, for example, one can generalize the
techniques in \cite[Proof of Th. 2]{cap_theorems:ShangKramerChen} for a fading
IFC with non-negative real $H_{j,k}$ for all $j,k$, such that $H_{1,1}%
>H_{2,1}$ and $H_{2,2}>H_{1,2}$, to outer bound the sum-rate by
\begin{equation}
\mathbb{E}\left[  C\left(  \frac{\left\vert H_{1,1}\right\vert ^{2}%
P_{1}\left(  \mathbf{H}\right)  }{1+\left\vert H_{1,2}\right\vert ^{2}%
P_{2}\left(  \mathbf{H}\right)  }\right)  +C\left(  \frac{\left\vert
H_{2,2}\right\vert ^{2}P_{1}\left(  \mathbf{H}\right)  }{1+\left\vert
H_{2,1}\right\vert ^{2}P_{2}\left(  \mathbf{H}\right)  }\right)  \right]  ,
\end{equation}
we require that $\rho_{k}\left(  \mathbf{H}\right)  $ and $\sigma_{k}%
^{2}\left(  \mathbf{H}\right)  $, for all $\mathbf{H}$, satisfy%
\begin{equation}
H_{1,1}H_{1,2}\left(  1+H_{2,1}^{2}P_{1}\left(  \mathbf{H}\right)  \right)
+H_{2,2}H_{2,1}\left(  1+H_{1,2}^{2}P_{2}\left(  \mathbf{H}\right)  \right)
\leq H_{1,1}H_{2,2}. \label{UW2_Cond}%
\end{equation}
This implies that for a given fading statistics, every choice of feasible
power policies $\underline{P}\left(  \mathbf{H}\right)  $ must satisfy the
condition in (\ref{UW2_Cond}). With the exception of a few trivial channel
models, the condition in (\ref{UW2_Cond}) cannot in general be satisfied by
all power policies. One approach is to extend the results on sum-capacity and
the related noisy interference condition for PGICs in \cite[Proof of Th.
3]{cap_theorems:Shang_03} to ergodic fading IFCs. Despite the fact that
ergodic fading channels are simply a weighted combination of parallel
sub-channels, extending the results in \cite[Proof of Th. 3]%
{cap_theorems:Shang_03} are not in general straightforward.
\end{remark}

\subsection{Uniformly Mixed IFC: Proof of Theorem \ref{Th_Mix}}

The proof of Theorem \ref{Th_UW2} follows directly from bounding the
sum-capacity a UM\ IFC by the sum-capacities of a UW one-sided IFC and a US
one-sided IFC that result from eliminating links one of the two interfering
links. Achievability follows from using the US coding scheme for the strong
user and the UW coding scheme for the weak user.

\subsection{Uniformly Weak IFC: Proof of Theorem \ref{Th_UW2}}

The proof of Theorem \ref{Th_UW2} follows directly from bounding the
sum-capacity a UW\ IFC by that of a UW one-sided IFC that results from
eliminating one of the interfering links (eliminating an interfering link can
only improve the capacity of the network). Since two complementary one-sided
IFCs can be obtained thus, we have two outer bounds on the sum-capacity of a
UW IFC denoted by $S^{\left(  w,1\right)  }\left(  \underline{P}\left(
\mathbf{H}\right)  \right)  \ $and $S^{\left(  w,2\right)  }\left(
\underline{P}\left(  \mathbf{H}\right)  \right)  $ in (\ref{SC_UW2}), where
$S^{\left(  w,1\right)  }\left(  \underline{P}\left(  \mathbf{H}\right)
\right)  $ and $S^{\left(  w,2\right)  }\left(  \underline{P}\left(
\mathbf{H}\right)  \right)  $ are the bounds for one-sided UW IFCs with
$H_{2,1}=0$ and $H_{1,2}=0$, respectively.

\subsection{Hybrid One-Sided IFC\textit{: }Proof of Theorem \ref{Th_Hyb}}

The bound in (\ref{HK1_SR}) can be obtained from the following code
construction: user $1$ encodes its message $w_{1}$ across all sub-channels by
constructing independent Gaussian codebooks for each sub-channel to transmit
the same message. On the other hand, user 2 transmits two messages $\left(
w_{2p},w_{2c}\right)  $ jointly across all sub-channels by constructing
independent Gaussian codebooks for each sub-channel to transmit the same
message pair. The messages $w_{2p}$ and $w_{2c}$ are transmitted at (fading
averaged) rates $R_{2p}$ and $R_{2c}$, respectively, such that $R_{2p}%
+R_{2c}=R_{2}$. Thus, across all sub-channels, one may view the encoding as a
Han Kobayashi coding scheme for a one-sided non-fading IFC in which the two
transmitted signals in each use of sub-channel $\mathbf{H}$ are
\begin{align}
X_{1}\left(  \mathbf{H}\right)   &  =\sqrt{P_{1}\left(  \mathbf{H}\right)
}V_{1}\left(  \mathbf{H}\right) \label{X1_hyb}\\
X_{2}\left(  \mathbf{H}\right)   &  =\sqrt{\alpha_{\mathbf{H}}P_{2}\left(
\mathbf{H}\right)  }V_{2}\left(  \mathbf{H}\right)  +\sqrt{\overline{\alpha
}_{\mathbf{H}}P_{2}\left(  \mathbf{H}\right)  }U_{2}\left(  \mathbf{H}\right)
\label{X2_hyb}%
\end{align}
where $V_{1}\left(  \mathbf{H}\right)  $, $V_{2}\left(  \mathbf{H}\right)  $,
and $U_{2}\left(  \mathbf{H}\right)  $ are independent zero-mean unit variance
Gaussian random variables, for all $\mathbf{H}$, $\alpha_{\mathbf{H}}%
\in\left[  0,1\right]  $ and $\overline{\alpha}_{\mathbf{H}}=1-\alpha
_{\mathbf{H}}$ are the power fractions allocated for $w_{2p}$ and $w_{2c}$,
respectively. Thus, over $n$ uses of the channel, $w_{2p}$ and $w_{2c}$ are
encoded via $V_{2}^{n}$ and $U_{2}^{n}$, respectively.

Receiver $1$ decodes $w_{1}$ and $w_{2c}$ jointly and receiver $2$ decodes
$w_{2p}$ and $w_{2c}$ jointly across all channel states provided
\begin{subequations}
\label{R2bounds}%
\begin{align}
R_{2p} &  \leq\mathbb{E}\left[  C\left(  \left\vert H_{2,2}\right\vert
^{2}\alpha_{\mathbf{H}}P_{2}\left(  \mathbf{H}\right)  \right)  \right]  \\
R_{2p}+R_{2c} &  \leq\mathbb{E}\left[  C\left(  \left\vert H_{2,2}\right\vert
^{2}P_{2}\left(  \mathbf{H}\right)  \right)  \right]
\end{align}
\end{subequations}
\begin{subequations}
\label{Hyb_HK}%
\begin{align}
R_{1} &  \leq\mathbb{E}\left[  C\left(  \frac{\left\vert H_{1,1}\right\vert
^{2}P_{1}\left(  \mathbf{H}\right)  }{1+\left\vert H_{1,2}\right\vert
^{2}\alpha_{\mathbf{H}}P_{2}\left(  \mathbf{H}\right)  }\right)  \right]  \\
R_{2c} &  \leq\mathbb{E}\left[  C\left(  \frac{\left\vert H_{1,2}\right\vert
^{2}\overline{\alpha}_{\mathbf{H}}P_{2}\left(  \mathbf{H}\right)
}{1+\left\vert H_{1,2}\right\vert ^{2}\alpha_{\mathbf{H}}P_{2}\left(
\mathbf{H}\right)  }\right)  \right]  \\
R_{1}+R_{2c} &  \leq\mathbb{E}\left[  C\left(  \frac{\left\vert H_{1,1}%
\right\vert ^{2}P_{1}\left(  \mathbf{H}\right)  +\left\vert H_{1,2}\right\vert
^{2}\overline{\alpha}_{\mathbf{H}}P_{2}\left(  \mathbf{H}\right)
}{1+\left\vert H_{1,2}\right\vert ^{2}\alpha_{\mathbf{H}}P_{2}\left(
\mathbf{H}\right)  }\right)  \right]  .
\end{align}
Using Fourier-Motzhkin elimination, we can simplify the bounds in
(\ref{R2bounds}) and (\ref{Hyb_HK}) to obtain
\end{subequations}
\begin{subequations}
\label{HybHKFin}%
\begin{align}
R_{1} &  \leq\mathbb{E}\left[  C\left(  \frac{\left\vert H_{1,1}\right\vert
^{2}P_{1}\left(  \mathbf{H}\right)  }{1+\left\vert H_{1,2}\right\vert
^{2}\alpha_{\mathbf{H}}P_{2}\left(  \mathbf{H}\right)  }\right)  \right]
\label{R1h}\\
R_{2} &  \leq\mathbb{E}\left[  C\left(  \left\vert H_{2,2}\right\vert
^{2}P_{2}\left(  \mathbf{H}\right)  \right)  \right]  \label{R2h}\\
R_{2} &  \leq\mathbb{E}\left[  C\left(  \alpha_{\mathbf{H}}\left\vert
H_{2,2}\right\vert ^{2}P_{2}\left(  \mathbf{H}\right)  \right)  \right]
+\mathbb{E}\left[  \frac{\left\vert H_{1,2}\right\vert ^{2}\overline{\alpha
}_{\mathbf{H}}P_{2}\left(  \mathbf{H}\right)  }{1+\left\vert H_{1,2}%
\right\vert ^{2}\alpha_{\mathbf{H}}P_{2}\left(  \mathbf{H}\right)  }\right]
\label{R2h2}\\
R_{1}+R_{2} &  \leq\mathbb{E}\left[  C\left(  \left\vert H_{2,2}\right\vert
^{2}\alpha_{\mathbf{H}}P_{2}\left(  \mathbf{H}\right)  \right)  \right]
+\mathbb{E}\left[  C\left(  \frac{\left\vert H_{1,1}\right\vert ^{2}%
P_{1}\left(  \mathbf{H}\right)  +\left\vert H_{1,2}\right\vert ^{2}%
\overline{\alpha}_{\mathbf{H}}P_{2}\left(  \mathbf{H}\right)  }{1+\left\vert
H_{1,2}\right\vert ^{2}\alpha_{\mathbf{H}}P_{2}\left(  \mathbf{H}\right)
}\right)  \right]  .\label{R12h}%
\end{align}
Combining the bounds in (\ref{HybHKFin}), for every choice of $\left(
\alpha_{\mathbf{H}},\underline{P}\left(  \mathbf{H}\right)  \right)  $, the
sum-rate is given by the minimum of two functions $S_{1}\left(  \alpha
_{\mathbf{H}},\underline{P}\left(  \mathbf{H}\right)  \right)  $ and
$S_{2}\left(  \alpha_{\mathbf{H}},\underline{P}\left(  \mathbf{H}\right)
\right)  $, where $S_{1}\left(  \underline{P}\left(  \mathbf{H}\right)
\right)  $ is the sum of the bounds on $R_{1}$ and $R_{2}$ in (\ref{R1h}) and
(\ref{R2h}), respectively, and $S_{2}\left(  \alpha_{\mathbf{H}},\underline
{P}\left(  \mathbf{H}\right)  \right)  $ is the bound on $R_{1}+R_{2}$ in
(\ref{R12h}). The bound on $R_{1}+R_{2}$ from combining (\ref{R1h}) and
(\ref{R2h2}) is at least as much as (\ref{R12h}), and hence, is ignored.

The maximization of the minimum of $S_{1}\left(  \underline{P}\left(
\mathbf{H}\right)  \right)  $ and $S_{2}\left(  \alpha_{\mathbf{H}}%
,\underline{P}\left(  \mathbf{H}\right)  \right)  $ can be shown to be
equivalent to a \textit{minimax }optimization problem (see for e.g.,
\cite[II.C]{cap_theorems:HVPoor01}) for which the maximum sum-rate $S^{\ast}$
is given by three cases. The three cases are defined below. Note that in each
case, the optimal $\underline{P}^{\ast}\left(  \mathbf{H}\right)  $ and
$\alpha_{\mathbf{H}}^{\ast}$ maximize the smaller of the two functions and
therefore maximize both in case when the two functions are equal. The three
cases are%

\end{subequations}
\begin{subequations}
\label{Hyb_MM}%
\begin{align}
&
\begin{array}
[c]{cc}%
\text{Case }1: & S^{\ast}=S_{1}\left(  \alpha_{\mathbf{H}}^{\ast}%
,\underline{P}^{\ast}\left(  \mathbf{H}\right)  \right)  <S_{2}\left(
\alpha_{\mathbf{H}}^{\ast},\underline{P}^{\ast}\left(  \mathbf{H}\right)
\right)
\end{array}
\label{MMC1}\\
&
\begin{array}
[c]{cc}%
\text{Case }2: & S^{\ast}=S_{2}\left(  \alpha_{\mathbf{H}}^{\ast}%
,\underline{P}^{\ast}\left(  \mathbf{H}\right)  \right)  <S_{1}\left(
\alpha_{\mathbf{H}}^{\ast},\underline{P}^{\ast}\left(  \mathbf{H}\right)
\right)
\end{array}
\label{MMC2}\\
&
\begin{array}
[c]{cc}%
\text{Case }3: & S^{\ast}=S_{1}\left(  \alpha_{\mathbf{H}}^{\ast}%
,\underline{P}^{\ast}\left(  \mathbf{H}\right)  \right)  =S_{2}\left(
\alpha_{\mathbf{H}}^{\ast},\underline{P}^{\ast}\left(  \mathbf{H}\right)
\right)
\end{array}
\label{MMC3}%
\end{align}
Thus, for Cases 1 and 2$,$ the minimax policy is the policy maximizing
$S_{1}\left(  \underline{P}\left(  \mathbf{H}\right)  \right)  $ and
$S_{2}\left(  \alpha_{\mathbf{H}},\underline{P}\left(  \mathbf{H}\right)
\right)  $ subject to the conditions in (\ref{MMC1}) and (\ref{MMC2}),
respectively, while for Case $3$, it is the policy maximizing $S_{1}\left(
\underline{P}\left(  \mathbf{H}\right)  \right)  $ subject to the equality
constraint in (\ref{MMC3}). We now consider this maximization problem for each
sub-class. Before proceeding, we observe that, $S_{1}\left(  \cdot\right)  $
is maximized for $\alpha_{\mathbf{H}}^{\ast}=0$ and $P_{k}^{\ast}\left(
\mathbf{H}\right)  =P_{k}^{(wf)}\left(  H_{kk}\right)  $, $k=1,2.$ On the
other hand, the $\alpha_{\mathbf{H}}^{\ast}$ maximizing $S_{2}\left(
\cdot\right)  $ depends on the sub-class.

\textit{Uniformly Strong}: The bound $S_{2}\left(  \alpha_{\mathbf{H}%
},\underline{P}\left(  \mathbf{H}\right)  \right)  $ in (\ref{R12h}) can be
rewritten as
\end{subequations}
\begin{equation}
\mathbb{E}\left[  C\left(  \left\vert H_{2,2}\right\vert ^{2}\alpha
_{\mathbf{H}}P_{2}\left(  \mathbf{H}\right)  \right)  \right]  -\mathbb{E}%
\left[  C\left(  \left\vert H_{1,2}\right\vert ^{2}\alpha_{\mathbf{H}}%
P_{2}\left(  \mathbf{H}\right)  \right)  \right]  +\mathbb{E}\left[  C\left(
\left\vert H_{1,1}\right\vert ^{2}P_{1}\left(  \mathbf{H}\right)  +\left\vert
H_{1,2}\right\vert ^{2}P_{2}\left(  \mathbf{H}\right)  \right)  \right]  ,
\label{US1}%
\end{equation}
and thus, when $\Pr[\left\vert H_{1,2}\right\vert >\left\vert H_{2,2}%
\right\vert ]=1$, for every choice of $\underline{P}\left(  \mathbf{H}\right)
$, $S_{2}\left(  \alpha_{\mathbf{H}},\underline{P}\left(  \mathbf{H}\right)
\right)  $ is maximized by $\alpha_{\mathbf{H}}=0$, i.e., $w_{2}=w_{2c}$. The
sum-capacity is given by (\ref{US_SC}) with $H_{2,1}=\infty$ (this is
equivalent to a genie aiding one of the receivers thereby simplifying the
sum-capacity expression in (\ref{US_SC}) for a two-sided IFC to that for a
one-sided IFC). Furthermore, $\alpha_{\mathbf{H}}=0$ also maximizes
$S_{1}\left(  \alpha_{\mathbf{H}},\underline{P}\left(  \mathbf{H}\right)
\right)  .$ In conjunction with the outer bounds for US IFCs developed
earlier, the US sum-capacity and the optimal policy achieving it are obtained
via the minimax optimization problem with $\alpha_{\mathbf{H}}^{\ast}=0$ such
that every sub-channel carries the same common information.

\textit{Uniformly Weak}: For this sub-class of channels, it is straightforward
to verify that for $\alpha_{\mathbf{H}}^{\ast}=0$ (\ref{MMC1}) will not be
satisfied. Thus, one is left with Cases 2 and $3$. From Theorem \ref{Th_UW1},
we have that $\alpha_{\mathbf{H}}^{\ast}=1$ achieves the sum-capacity of
one-sided UW\ IFCs, i.e., $w_{2}=w_{2p}$. Furthermore, $S_{2}\left(
1,\underline{P}\left(  \mathbf{H}\right)  \right)  =S_{1}\left(
1,\underline{P}\left(  \mathbf{H}\right)  \right)  $, and thus, the condition
for Case $2$ is not satisfied, i.e., this sub-class corresponds to Case 3 in
the minimax optimization. The constrained optimization in (\ref{MMC3}) for
Case $3$ can be solved using Lagrange multipliers though the solution is
relatively easier to develop using techniques in Theorem \ref{Th_UW1}.

\textit{Ergodic Very Strong}: As mentioned before, $S_{1}\left(  \cdot\right)
$ is maximized for $\alpha_{\mathbf{H}}^{\ast}=0$ and $P_{k}^{\ast}\left(
\mathbf{H}\right)  =P_{k}^{(wf)}\left(  H_{kk}\right)  $, $k=1,2$, i.e. when
$w_{2}=w_{2c}$ and each user waterfills on its intended link. From
(\ref{Hyb_MM})$,$ we see that the sum-capacity of EVS IFCs is achieved
provided the condition for Case 1 in (\ref{Hyb_MM}) is satisfied. Note that
this maximization does not require the sub-channels to be UW or US.

\textit{Hybrid}: When the condition for Case $1$ in (\ref{Hyb_MM}) with
$\alpha_{\mathbf{H}}^{\ast}=0$ is satisfied, we obtain an EVS\ IFC. On the
other hand, when this condition is not satisfied, the optimization simplifies
to considering Cases $2$ and 3, i.e., $\alpha_{\mathbf{H}}^{\ast}\not =0$ for
all $\mathbf{H}$. Using the linearity of expectation, we can write the
expressions for $S_{1}\left(  \cdot\right)  $ and $S_{2}\left(  \cdot\right)
$ as sums of expectations of the appropriate bounds over the collection of
weak and strong sub-channels. Let $S_{k}^{\left(  w\right)  }\left(
\cdot\right)  $ and $S_{k}^{\left(  s\right)  }\left(  \cdot\right)  $ denote
the expectation over the weak and strong sub-channels, respectively, for
$k=1,2$, such that $S_{k}\left(  \cdot\right)  =S_{k}^{\left(  w\right)
}\left(  \cdot\right)  +S_{k}^{\left(  s\right)  }\left(  \cdot\right)  $,
$k=1,2.$

Consider Case $2$ first. For those sub-channels which are strong, one can use
(\ref{US1}) to show that $\alpha_{\mathbf{H}}^{\ast}=0$ maximizes
$S_{2}^{\left(  s\right)  }\left(  \cdot\right)  $. Suppose we choose
$\alpha_{\mathbf{H}}^{\ast}=1$ to maximize $S_{2}^{\left(  w\right)  }\left(
\cdot\right)  $. From the UW analysis earlier, $S_{2}^{\left(  w\right)
}\left(  1,P\left(  \mathbf{H}\right)  \right)  =S_{1}^{\left(  w\right)
}\left(  1,P\left(  \mathbf{H}\right)  \right)  $, and therefore, (\ref{MMC2})
is satisfied only when $S_{2}^{\left(  s\right)  }\left(  0,P\left(
\mathbf{H}\right)  \right)  <S_{1}^{\left(  s\right)  }\left(  0,P\left(
\mathbf{H}\right)  \right)  $. This requirement may not hold in general, and
thus, to satisfy (\ref{MMC2}), we require that $\alpha_{\mathbf{H}}^{\ast}%
\in(0,1]$ for those $\mathbf{H}$ that represent weak sub-channels. Similar
arguments hold for Case $3$ too thereby justifying (\ref{alpstar_hyb}) in
Theorem \ref{Th_Hyb}.

\begin{remark}
The bounds in (\ref{R2bounds}) are written assuming superposition coding of
the common and private messages at transmitter $2$. The resulting bounds
following Fourier-Motzkin elimination remain unchanged even if we included an
additional bound on $R_{2c}$ at receiver $2$ in (\ref{R2bounds}).
\end{remark}

\section{\label{Sec_Dis}Discussion}

\begin{figure}[tbp] \centering
{\includegraphics[
height=2.8271in,
width=5.6213in
]%
{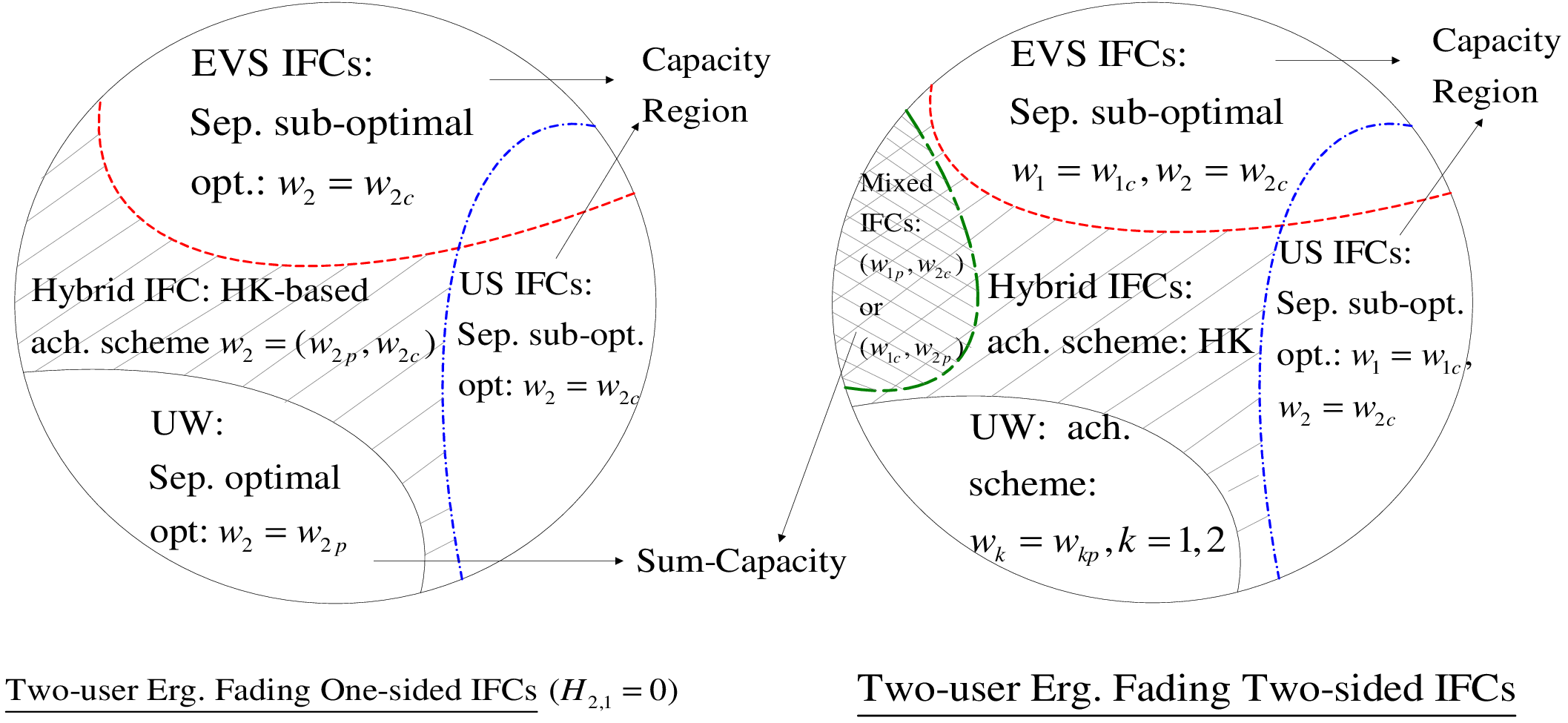}%
}%
\caption{Overview of capacity results for two-sided and one-sided ergodic fading IFCs.}\label{Fig_IFCVenn}%
\end{figure}

As in the non-fading case (see \cite{cap_theorems:ETW} for a detailed
development of outer bounds), the outer bounds and capacity results we have
obtained are in general tailored to specific regimes of fading statistics. Our
results can be summarized by two Venn diagrams, one for the two-sided and one
for the one-sided, as shown in Fig. \ref{Fig_IFCVenn}. Taking a Han-Kobayashi
view-point, the diagrams show that transmitting common messages is optimal for
the EVS and US IFCs, i.e., $w_{k}=w_{kc}$, $k=1,2$. Similarly, choosing only a
private message at the interfering transmitter, i.e., $w_{2}=w_{2p}$ for
$H_{2,1}=0$ and $w_{1}=w_{1p}$ for $H_{1,2}=0$, is optimal for the one-sided
UW\ IFC. For the mixed IFCs, it is optimal for the strongly and the weakly
interfering users to transmit only common and only private messages,
respectively. For the remaining hybrid IFCs and two-sided UW IFCs, the most
general achievable strategy results from generalizing the HK scheme to the
fading model, i.e., each transmitter in the two-sided IFC transmits private
and common messages while only the interfering transmitter does so in the
one-sided model. These results are summarizes in Fig. \ref{Fig_IFCVenn}. The
sub-classes for which either the sum-capacity or the entire capacity region is
known are also indicated in the Figure.%

We now present examples of continuous and discrete fading process for which
the channel states satisfy the EVS condition. Without loss of generality in
both examples we assume that the direct links are non-fading. Thus, for the
case where the fading statistics and average power constraints $\overline
{P}_{k}$ satisfy the EVS\ conditions in (\ref{EVS_Cond}), it is optimal for
transmitter $k$ to transmit at $\overline{P}_{k}$. For the continuous model,
we assume that the cross-links are independent and identically distributed
Rayleigh faded links, i.e., $H_{j,k}\sim\mathcal{CN}\left(  0,\sigma
^{2}/2\right)  $ for all $j\not =k,j,k=1,2$. For the discrete model, we assume
that the cross-link fading states take values in a binary set $\left\{
h_{1},h_{2}\right\}  $. Finally, we set $\overline{P}_{1}=\overline{P}%
_{2}=\overline{P}$.

\begin{figure*}[tbp] \centering
{\includegraphics[
trim=0.167262in 0.052738in 0.216967in 0.064759in,
height=3.154in,
width=5.9931in
]%
{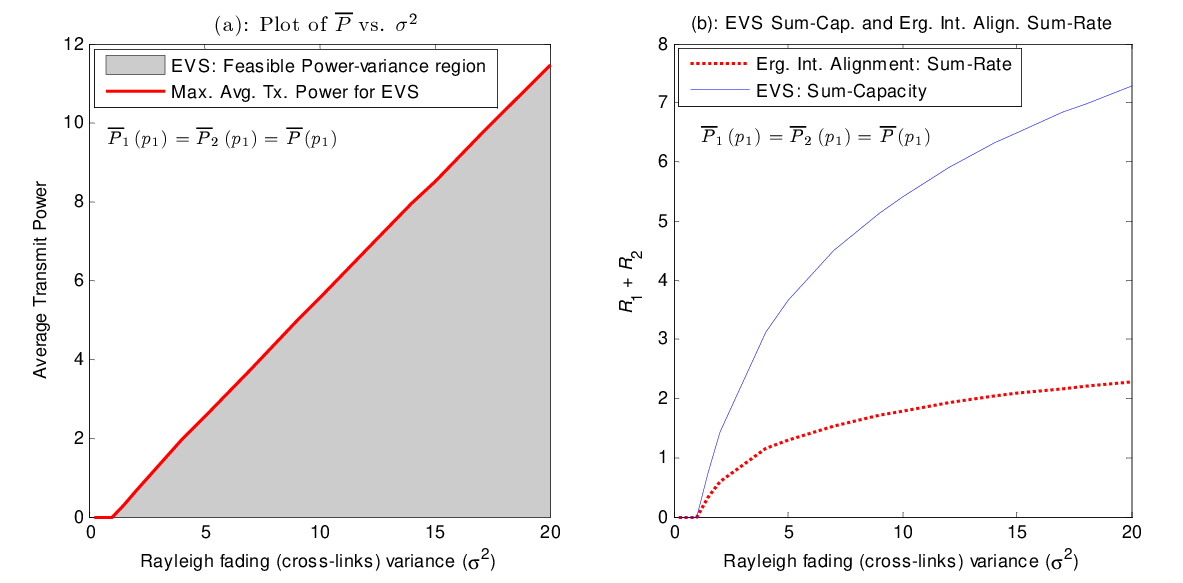}%
}%
\caption{Feasible Power-variance region for EVS, EVS sum-capacity, and Ergodic Interference Alignment Sum-Rate.}\label{Fig_RayLIFC}%
\end{figure*}

For every choice of the Rayleigh fading variance $\sigma^{2}$, we determine
the maximum $\overline{P}$ for which the EVS conditions in (\ref{EVS_Cond})
hold. The resulting feasible $\overline{P}$ vs. $\sigma^{2}$ region is plotted
in\ Fig. \ref{Fig_RayLIFC}(a). Our numerical results indicate that for very
small values of $\sigma^{2}$, i.e., $\sigma^{2}<1.5$, where the cumulative
distribution of fading states with $\left\vert H_{j,k}\right\vert \,<1$ is
close to $1$, the EVS condition cannot be satisfied by any finite value of
$\overline{P}$, however small. As $\sigma^{2}$ increases thereby increasing
the likelihood of $\left\vert H_{j,k}\right\vert \,>1$, $\overline{P}$
increases too. Also plotted in Fig. \ref{Fig_RayLIFC}(b) is the EVS
sum-capacity achieved at $\overline{P}_{\max}$, the maximum $\overline{P}$ for
every choice of $\sigma^{2}$. Furthermore, since the Rayleigh fading channel
allows ergodic interference alignment \cite{cap_theorems:Nazer01}, we compare
the EVS\ sum-capacity with the sum-rate achieved by ergodic interference
alignment for every choice of $\sigma^{2}$ and the corresponding $\overline
{P}_{\max}$. This achievable scheme, whose sum-rate is the same as that
achieved when the users are time-duplexed, is closer to the sum-capacity only
for small values of $\sigma^{2}$. This is to be expected as EVS\ IFCs achieve
the largest possible degrees of freedom, which is $2$ for a two-user IFC while
the scheme of achieves at most one degree of freedom.

From (\ref{EVS_Cond}), one can verify that for a non-fading very strong IFC,
the very strong condition sets an upper bound on the average transmit power
$\overline{P}_{k}$ at user $k$ as%
\begin{equation}%
\begin{array}
[c]{cc}%
\overline{P}_{k}<H_{k,j}/\left(  \left\vert H_{1,1}\right\vert ^{2}\left\vert
H_{2,2}\right\vert ^{2}\right)  -1 & j\not =k,j,k\in\left\{  1,2\right\}  .
\end{array}
\end{equation}
One can view the upper bound on $\overline{P}$ for the EVS\ IFCs in\ Fig.
\ref{Fig_RayLIFC} as an equivalent fading-averaged bound.%

\begin{figure}
[ptb]
\begin{center}
\includegraphics[
trim=0.258254in 0.038821in 0.399347in 0.150780in,
height=3.3217in,
width=4.3785in
]%
{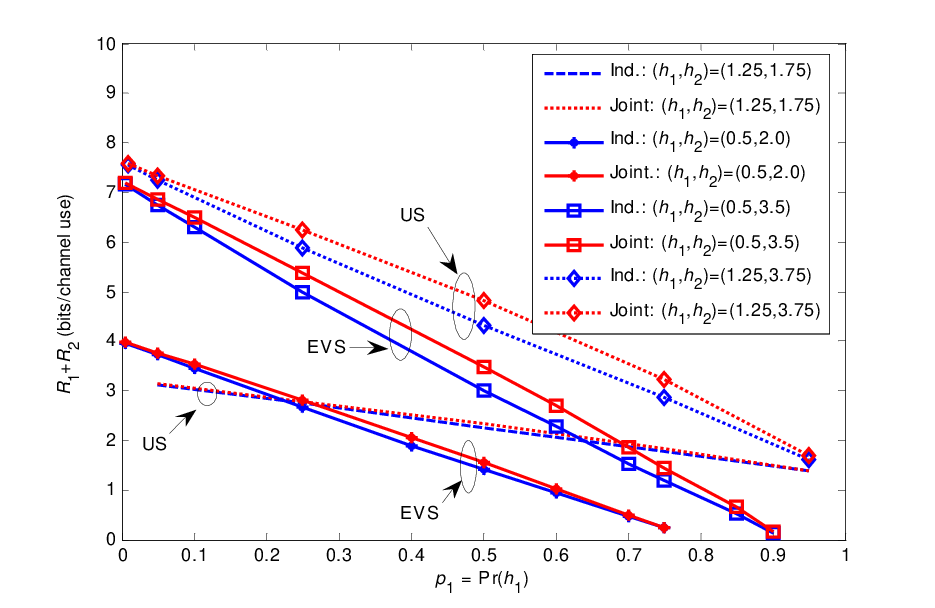}%
\caption{Plot comparing the sum-capacities and the sum-rates achieved by
separable coding for different values of $\left(  h_{1},h_{2}\right)  $ that
result in either an EVS\ or a US IFC.}%
\label{Fig_IFCSep}%
\end{center}
\end{figure}

We next compare the effect of joint and separate coding for one-sided EVS and
US IFCs. For computational simplicity, we consider a discrete fading model
where the non-zero cross-link fading state take values in a binary set
$\left\{  h_{1},h_{2}\right\}  $ while the direct links are non-fading unit
gains. For a one-sided EVS\ IFC, we choose $\left(  h_{1},h_{2}\right)
=\left(  0.5,3.5\right)  $ and $\overline{P}_{1}=\overline{P}_{2}=\overline
{P}_{\max}$ where $\overline{P}_{\max}$ is the maximum power for which the EVS
conditions in (\ref{EVS_Cond}) are satisfied (note that only one of the
conditions are relevant since it is a one-sided IFC). In Fig. \ref{Fig_IFCSep}%
, the EVS sum-capacity is plotted along with the sum-rate achieved by
independent coding in each sub-channel as a function of the probability
$p_{1}$ of the fading state $h_{1}$. Here independent coding means that each
sub-channel is viewed as a non-fading one-sided IFC and the sum-capacity
achieving strategy for each sub-channel is applied.

As expected, as $p_{1}\rightarrow0$ or $p_{1}\rightarrow1$, the sum-rate
achieved by separable coding approaches the joint coding scheme. Thus, the
difference between the optimal joint coding and the sub-optimal independent
coding schemes is the largest when both fading states are equally likely. In
contrast to this example where the gains from joint coding are not negligible,
we also plot in Fig. \ref{Fig_IFCSep} the sum-capacity and sum-rate achieved
by independent coding for an EVS\ IFC with $\left(  h_{1},h_{2}\right)
=\left(  0.5,2.0\right)  $ for which the rate difference is very small. Thus,
as expected, joint coding is advantageous when the variance of the cross-link
fading is large and the transmit powers are small enough to result in an EVS
IFC. In the same plot, we also compare the sum-capacity with the sum-rate
achieved by a separable scheme for two US IFCs, one given by $\left(
h_{1},h_{2}\right)  =\left(  1.25,1.75\right)  $ and the other by $\left(
h_{1},h_{2}\right)  =\left(  1.25,3.75\right)  $. As with the EVS\ examples,
here too, the rate difference between the optimal joint strategy and the, in
general, sub-optimal independent strategy increases with increasing variance
of the fading distribution.

One can similarly compare the performance of independent and joint coding for
two-sided EVS and US\ IFCs. In this case, the more general HK\ scheme needs to
be considered in each sub-channel for the independent coding case. In general,
the observations for the one-sided also extend to the two-sided IFC.%

Finally, we demonstrate sum-rates achievable by Theorem \ref{Th_Hyb} for a
hybrid one-sided IFC. As before, for computational simplicity, we consider a
discrete fading model where the cross-link fading states take values in a
binary set $\left\{  h_{1},h_{2}\right\}  $ while the direct links are
non-fading unit gains. Without loss of generality, we choose $\left(
h_{1},h_{2}\right)  =\left(  0.5,2.0\right)  $ and assume $\overline{P}%
_{1}=\overline{P}_{2}=\overline{P}$. The sum-rate achieved by the proposed
HK-like scheme, denoted $R_{sum}^{\left(  HK\right)  }$, is determined as a
function of the probability $p_{1}$ of the weak state $h_{1}$. For each
$p_{1}$, using the fact that a hybrid IFC is by definition one for which the
EVS condition is not satisfied, we choose $\overline{P}\left(  p_{1}\right)
=\overline{P}_{\max}^{EVS}\left(  p_{1}\right)  +1.5$ where $\overline
{P}_{\max}^{EVS}\left(  p_{1}\right)  $ is the maximum $\overline{P}$ for
which the EVS conditions hold for the chosen $p_{1}$ and $\left(  h_{1}%
,h_{2}\right)  $.

\begin{figure*}[tbp] \centering
{\includegraphics[
trim=0.180674in 0.000000in 0.241425in 0.000000in,
height=3.1938in,
width=6.1289in
]%
{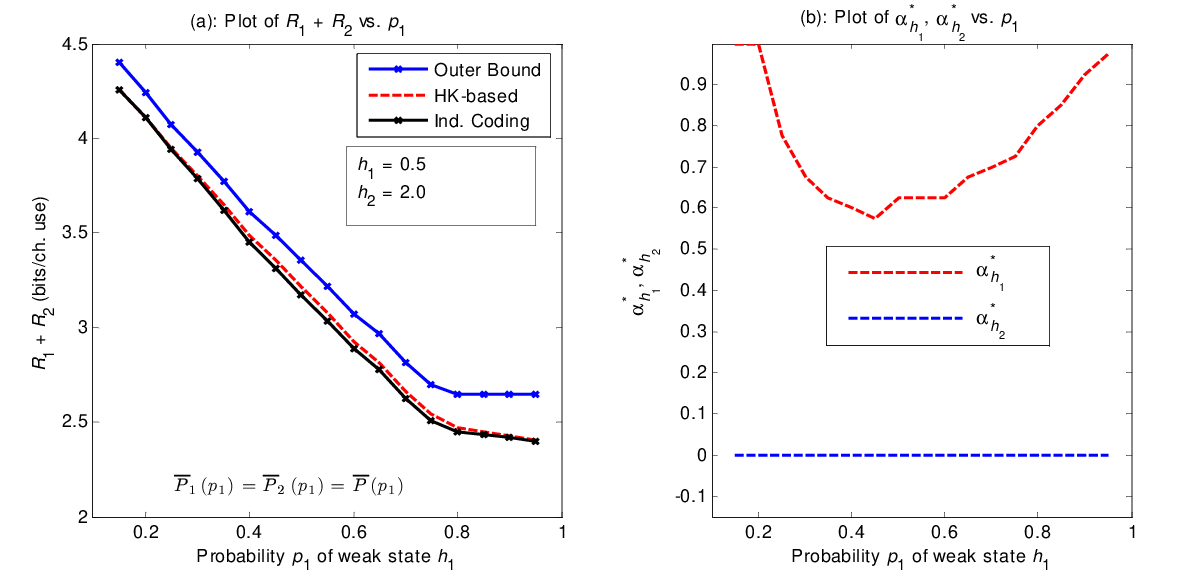}%
}%
\caption{Sum-Rate vs. $p_1$ for HK-based scheme and Separable coding scheme and plots of optimal power fractions for HK-based scheme.}\label{Fig_HKParOB}%
\end{figure*}

In Fig. \ref{Fig_HKParOB}(a), we plot $R_{sum}^{\left(  HK\right)  }$ as a
function of $p_{1}$. We also plot the largest sum-rate outer bounds
$R_{sum}^{\left(  OB\right)  }$ obtained by assuming interference-free links
from the users to the receivers. Finally, for comparison, we plot the sum-rate
$R_{sum}^{\left(  Ind\right)  }$ achieved by a separable coding scheme in each
sub-channel. This separable coding scheme is simply a special case of the
HK-based joint coding scheme presented for hybrid one-sided IFCs in Theorem
\ref{Th_Hyb} obtained by choosing $\alpha_{H}^{\ast}=0$ and $\alpha_{H}^{\ast
}=1$ in the strong and weak sub-channels, respectively. Thus, $R_{sum}%
^{\left(  Ind\right)  }\leq R_{sum}^{\left(  HK\right)  }$ as demonstrated in
the plot. In Fig. \ref{Fig_HKParOB}(b), the fractions $\alpha_{h_{1}}^{\ast}$
and $\alpha_{h_{2}}^{\ast}$ in the $h_{1}$ (weak) and the $h_{2}$ (strong)
states, respectively, are plotted. As expected, $\alpha_{h_{2}}^{\ast}=0$; on
the other hand, $\alpha_{h_{1}}^{\ast}$ varies between $0$ and $1$ such that
for $p_{1}\rightarrow1$, $\alpha_{h_{1}}^{\ast}\rightarrow1$ and for
$p_{1}\rightarrow1$, $\alpha_{h_{1}}^{\ast}\rightarrow1.$ Thus, when either
the weak or the strong state is dominant, the performance of the HK-based
coding\ scheme approaches that of the separable scheme in
\cite{cap_theorems:SumCap_Par_ZIFC}.%

\section{\label{Sec_Con}Conclusions}

We have developed the sum-capacity of specific sub-classes of ergodic fading
IFCs. These sub-classes include the ergodic very strong (mixture of weak and
strong sub-channels satisfying the EVS condition), the uniformly strong
(collection of strong sub-channels), the uniformly weak one-sided (collection
of weak one-sided sub-channels) IFCs, and the uniformly mixed (mix of UW and
US one-sided IFCs) two-sided IFCs. Specifically, we have shown that requiring
both receivers to decode both messages, i.e., simplifying the IFC to a
compound MAC, achieves the sum-capacity and the capacity region of the EVS and
US (one- and two-sided) IFCs. For both sub-classes, achieving the sum-capacity
requires encoding and decoding jointly across all sub-channels.

In contrast, for the UW one-sided IFCs, we have used genie-aided methods to
show that the sum-capacity is achieved by ignoring interference at the
interfered receiver and with independent coding across sub-channels. This
approach also allowed us to develop outer bounds on the two-sided UW\ IFCs. We
combined the UW and US one-sided IFCs results to develop the sum-capacity for
the uniformly mixed two-sided IFCs and showed that joint coding is optimal.

For the final sub-class of hybrid one-sided IFCs with a mix of weak and strong
sub-channels that do not satisfy the EVS conditions, using the fact that the
strong sub-channels can be exploited, we have proposed a Han-Kobayashi based
achievable scheme that allows partial interference cancellation using a joint
coding scheme. Assuming no time-sharing, we have shown that the sum-rate is
maximized by transmitting only a common message on the strong sub-channels and
transmitting a private message in addition to this common message in the weak
sub-channels. Proving the optimality of this scheme for the hybrid sub-class
remains open. However, we have also shown that the proposed joint coding
scheme applies to all sub-classes of one-sided IFCs, and therefore,
encompasses the sum-capacity achieving schemes for the EVS, US, and UW sub-classes.

Analogously with the non-fading IFCs, the ergodic capacity of a two-sided IFC
continues to remain unknown in general. However, additional complexity arises
from the fact that the sub-channels can in general be a mix of weak and strong
IFCs. A direct result of this complexity is that, in contrast to the
non-fading case, the sum-capacity of a one-sided fading IFC remains open for
the hybrid sub-class. The problem similarly remains open for the two-sided
fading IFC. An additional challenge for the two-sided IFC is that of
developing tighter bounds for the uniformly weak channel.

\bibliographystyle{IEEEtran}
\bibliography{IC_refs}

\end{document}